\documentclass[aps,prd,superscriptaddress,twocolumn,nofootinbib,preprintnumbers]{revtex4}
\usepackage{graphicx}
\usepackage{dcolumn}
\usepackage{bm}
\usepackage{natbib}
\usepackage{amsmath}
\usepackage{amssymb}
\usepackage{epstopdf}
\usepackage{hhline}
\usepackage{color}
\usepackage{relsize}
 \usepackage[titletoc,toc,title]{appendix}
\usepackage{array}
\newcolumntype{P}[1]{>{\centering\arraybackslash}p{#1}}
\usepackage{multirow}

\makeatletter

\newcommand{\Rmnum}[1]{\expandafter\@slowromancap\romannumeral #1@}
\makeatother

\begin{document} 
\title{Early-Universe Constraints on Dark Matter-Baryon Scattering and their Implications for a Global 21cm Signal}
\preprint{MIT/CTP-4995}

\author{Tracy R. Slatyer}
\email{tslatyer@mit.edu}
\affiliation{Center for Theoretical Physics, Massachusetts Institute of Technology, Cambridge, MA 02139, USA}

\author{Chih-Liang Wu}
\email{cliang@mit.edu}
\affiliation{Center for Theoretical Physics, Massachusetts Institute of Technology, Cambridge, MA 02139, USA}

\begin{abstract} 
We present and compare several cosmological constraints on the cross section for elastic scattering between dark matter (DM) and baryons, for cross sections with a range of power-law dependences on the DM-baryon relative velocity $v$, especially focusing on the case of $\sigma \propto v^{-4}$. We study constraints spanning a wide range of epochs in cosmological history, from pre-recombination distortions to the blackbody spectrum and anisotropies of the cosmic microwave background (CMB), to modifications to the intergalactic medium temperature and the resulting 21cm signal, and discuss the allowed signals in the latter channels given the constraints from the former. We improve previous constraints on DM-baryon scattering from the CMB anisotropies, demonstrate via principal component analysis that the effect on the CMB can be written as a simple function of DM mass for $v^{-4}$ scattering, and map out the redshifts dominating this signal. We show that given high-redshift constraints on DM-baryon scattering, a $v^{-4}$ scaling of the cross section for light DM would be sufficient to explain the deep 21cm absorption trough recently claimed by the EDGES experiment, if 100\% of the DM scatters with baryons. For millicharged DM models proposed to explain the observation, where only a small fraction of the DM interacts, we estimate that a PIXIE-like future experiment measuring CMB spectral distortion could test the relevant parameter space.
\end{abstract}

\maketitle

\section{Introduction}

Scattering between the visible matter and its dark-matter (DM) counterpart could potentially imprint a wide range of modifications on observational probes of the early universe. In particular, DM-proton or DM-electron scattering prior to recombination will generically (1) modify the evolving small-scale perturbations, leaving imprints on the temperature and polarization anisotropies of the cosmic microwave background (CMB) \cite{Dvorkin:2013cea, Gluscevic:2017ywp, Boddy:2018kfv, Xu:2018efh}, and (2) slightly cool or heat the visible matter, resulting in a spectral distortion to the CMB blackbody spectrum \cite{Ali-Haimoud:2015pwa}. At later times, the same heating or cooling processes could modify the thermal evolution of the baryonic matter, and in turn alter the absorption or emission of 21cm photons. After reionization, the Lyman-alpha forest provides an even lower-redshift probe of the gas temperature.

This possibility has recently garnered great interest due to the claimed detection of an enhanced 21cm absorption feature from $z\sim 15-20$, by the EDGES experiment \cite{bowman2018absorption}. It was quickly pointed out that such enhanced absorption could be due to a lower-than-expected gas temperature, which in turn could originate from DM-baryon scattering \cite{barkana2018possible}. In order to evade other constraints, it has been proposed that the DM-baryon scattering cross section could have a large velocity dependence, e.g. $\sigma \propto v^{-4}$. This would lead to a strong signal at the end of the cosmic dark ages, while suppressing scattering at large redshifts where thermal velocities are higher, as well as within Galactic halos where gravitational potential wells increase the velocity dispersion.

Several authors \cite{Munoz:2018pzp, Berlin:2018sjs, Barkana:2018qrx} have considered the possibility of DM possessing some small electric millicharge, which would ensure a cross-section of this $v^{-4}$ form from the scattering due to the Coulomb interaction (analogous to Rutherford scattering). In this case, only a small fraction of the DM should be millicharged, in order to evade constraints from the early universe and the distribution of DM in the present day. These works have identified a nominally allowed region of parameter space, capable of producing a low enough gas temperature to generate the EDGES absorption signal, for 10-80 MeV millicharged DM, comprising $0.3-2\%$ of the DM, and with a millicharge in the $10^{-4}-10^{-6}$ range.

We note that there are other possible explanations for the EDGES claim, even if foreground and instrumental effects are excluded:
\begin{itemize}
\item Additional radiation backgrounds in the relevant frequency range could enhance 21cm absorption \cite{Feng:2018rje}, in lieu of a low gas temperature. These backgrounds could potentially originate either from a DM-related source \cite{Fraser:2018acy,Pospelov:2018kdh} or from astrophysical objects such as black holes (e.g. \cite{Bernal:2017nec,Ewall-Wice:2018bzf}).
\item A lowered gas temperature could in principle be achieved by a mechanism other than scattering off a colder thermal bath; e.g. any phenomenon that causes the baryons and CMB to decouple earlier than expected will lower the late-time gas temperature \cite{hongwantoappear}.
\end{itemize}
Finally, the DM-baryon scattering cross section could be enhanced at low redshift / suppressed at high redshift by mechanisms other than velocity dependence; for example, if the scattering component is absent at early times but produced at late times through decays or oscillations.

In this work, however, we will focus on adapting, understanding, and extending the existing cosmological constraints on velocity-dependent scattering, and testing their compatibility with the EDGES signal. We leave other directions for future studies.

We present constraints on DM-baryon scattering from measurements of the CMB anisotropy spectrum, following Ref.~\cite{Xu:2018efh}, for the cases of $\sigma \propto v^n$ with $n=0$ and $n=-4$. We study the effect on the constraints of adding high-$\ell$ data from ACT and SPT, and find a modest improvement in the limits for $n=-4$, with a more pronounced improvement for $n=0$. We determine which redshifts contribute most strongly to the signal for $n=0$ and $n=-4$, by considering turning on scattering for limited redshift ranges, and validate a Fisher-matrix approach to estimating the detectability of scattering. We perform a principal component analysis and confirm the suggestion of Ref.~\cite{Xu:2018efh} that the mass dependence of the constraints can be parameterized in a simple way for $n=-4$ scattering; we quantitatively estimate the error in this parameterization to be at the percent level. Such a parameterization is also valid for $n=0$ scattering at masses above $\sim 0.1$ GeV, but breaks down at low masses, as we will discuss.\footnote{We thank Vera Gluscevic for valuable discussions which clarified our understanding of this point.} Again using principal component analysis, we provide a basis of redshift-dependent scattering histories with orthogonal effects on the CMB (after marginalization over the other cosmological parameters), which can be used to estimate constraints on general scattering histories.

Assuming that 100\% of the DM scatters on baryons with the given cross section, we compute the maximum modification to the low-$z$ gas temperature consistent with the cosmological constraints for $n=0$ and $n=-4$. We find that at $z=17$, the maximum temperature decrease for $n=0$ scattering is below $10^{-3}$ K, and can thus safely be neglected; however, for $n=-4$ scattering, changes in the gas temperature of several K at $z\sim 17$ can indeed be consistent with the CMB constraints -- at least within our current approximations -- for light DM with mass below 1 GeV.

We then examine constraints on DM-baryon scattering from spectral distortions of the CMB blackbody, following the methodology of \cite{Ali-Haimoud:2015pwa} (which considered scattering with $n \ge -2$). We go beyond the approximations of \cite{Ali-Haimoud:2015pwa} in order to estimate constraints from FIRAS and the sensitivity of a future PIXIE-like experiment for $n=-4$ scattering.

While these constraints are nominally weaker than those from the CMB anisotropies, they measure the energy losses from the photon-baryon fluid due to scattering with DM, and so can still provide non-negligible constraints on a small fraction of the DM interacting with baryons (or photons), in the same way that a small fraction of the DM interacting with the gas could cool the gas at late times. In the regime where the perturbation to the DM temperature for the interacting component is small, the fraction of DM that interacts is degenerate with the DM-baryon scattering cross section. Thus this bound can be used to constrain the scenario of a subdominant millicharged component, in contrast to the constraints from the CMB anisotropies, which we expect to become invalid if the scattering component is too small. In particular, if the scattering component is smaller than the uncertainties in the DM and baryon abundances, it is difficult to see how it could be constrained by the CMB; for very large cross sections leading to tight coupling between this scattering component and the baryons, it could appear simply as a slight increase in the overall baryon abundance \cite{Berlin:2018sjs}.

Furthermore, if the scattering component is indeed millicharged, its scatterings with visible matter at $z\sim 17$ are only with the small ionized fraction of the gas, $x_e \sim 2\times 10^{-4}$, whereas at redshifts prior to recombination relevant for spectral distortions ($10^3 < z < 10^6$), the ionization fraction is close to 1. Consequently, we estimate that near-future experiments could have sensitivity to the region of parameter space relevant to EDGES, in the scenario where a small fraction of DM is millicharged.

Finally, for completeness, we review constraints from the gas temperature after reionization, and compute new limits for the case of $n=-2$; however, these constraints are in general weaker than the others we consider.

The paper is organized as follows. In Sec.~\ref{sec:formalism}, we review the formalism to compute the modified evolution of cosmological perturbations and temperature in the presence of DM-baryon scattering, in particular for scattering cross sections with a power-law dependence on the relative velocity. In Sec.~\ref{sec:cmbconstraints}, we revisit, explore and extend the limits on DM-baryon scattering from CMB anisotropies. In Sec.~\ref{sec:heating} we discuss the implications of these constraints for cooling of the inter-galactic medium (IGM) by DM-baryon scattering during the cosmic dark ages, and also review limits from changes to the IGM temperature after reionization. In Sec.~\ref{sec:distortion} we discuss spectral distortions to the CMB from DM-baryon scattering, and extend previous constraints to the case of $\sigma \propto v^{-4}$ and models where a small fraction of DM carries electric millicharge. We conclude the paper in Sec.~\ref{sec:conclusion}.

\section{Formalism}
\label{sec:formalism}
We briefly review the effects of DM-baryon scattering on temperature and perturbation evolution in this section (for further details see Ref.~\cite{Dvorkin:2013cea}). For each Fourier mode with wavenumber $k$, we solve the evolution equations for the DM (denoted $\chi$) and baryon (denoted $b$) density fluctuations ($\delta_\chi$ and $\delta_b$) and velocity divergences ($\theta_\chi$ and $\theta_b$). We work in synchronous gauge, but introduce the velocity divergence $\theta_\chi$ that represents a nonzero peculiar velocity for DM, arising from the interaction with baryons.  We must also take into account the evolution of the DM and baryon temperatures, denoted $T_\chi$ and $T_b$ respectively; when the DM is light and the self-interaction cross section is substantial then $T_\chi$ can be non-negligible.

The CMB power spectra in the presence of DM-baryon interactions are governed by the following sets of equations \cite{Sigurdson:2004zp}:
\begin{eqnarray}
\dot{\delta_\chi} &=& - \theta_\chi-\dfrac{\dot{h}}{2}, \nonumber \\
\dot{\delta_b} &=& - \theta_b-\dfrac{\dot{h}}{2}, \nonumber \\
\dot{\theta_\chi} &=& - \dfrac{\dot{a}}{a}\theta_\chi+c_\chi^2k^2\delta_\chi + R_\chi \left(\theta_b-\theta_\chi \right), \nonumber \\
\dot{\theta_b} &=& - \dfrac{\dot{a}}{a}\theta_b+c_b^2 k^2 \delta_b + R_\gamma \left(\theta_\gamma-\theta_b \right) \nonumber \\
&& +\dfrac{\rho_\chi}{\rho_b}R_\chi \left(\theta_\chi-\theta_b \right), \nonumber \\
\dot{\theta_\gamma} & = & k^2 \left( \frac{1}{4} \delta_\gamma -\sigma_\gamma \right) - \frac{1}{\tau_c} (\theta_\gamma - \theta_b).
\end{eqnarray}
where $c_\chi$ and $c_b$ are the sound speeds (for DM/baryons respectively) defined by:
\begin{eqnarray}
c_b^2 &=&  \dfrac{k_B T_b}{\mu_b} \left( 1- \dfrac{1}{3} \dfrac{d \, \text{ln}  \,T_b}{d  \, \text{ln} \,a} \right),\nonumber \\
c_\chi^2 &=&  \dfrac{k_B T_\chi}{m_\chi} \left( 1- \dfrac{1}{3} \dfrac{d  \,\text{ln}  \,T_\chi}{d \, \text{ln} \,a} \right),
\end{eqnarray}
In these equations, $h$ is the trace of the metric perturbation, $\sigma_\gamma$ describes the shear stress, $\tau_c^{-1} = a n_e \sigma_T$, $R_\gamma$ is the Compton collision rate given by $R_\gamma = (4/3) (\rho_\gamma/\rho_b) a n_e \sigma_T$, and $R_\chi$ is the DM-baryon momentum exchange rate. $n_e$ is the free electron density and $\sigma_T$ is the Thomson cross section, and $\rho_X$ ($T_X$) for some species $X$ denotes its energy density (temperature).

Suppose the scattering cross section has a power-law dependence on redshift:
\begin{equation} \sigma = \sigma_0 v^n.\end{equation}
Then if we neglect for the moment any bulk relative velocity between the DM and baryon fluids, which is a good approximation for redshifts $z > 10^4$, and considering only scattering on hydrogen and helium,
$R_\chi$ can be written as \cite{Xu:2018efh}:
\begin{align}
 R_\chi & = \dfrac{a c_n \rho_b \sigma_0 }{m_\chi+m_H} \left( \dfrac{T_b}{m_H}+\dfrac{T_\chi}{m_\chi}\right)^{\frac{n+1}{2}}F_\text{He},
\end{align}
where the numerical prefactor $c_n$ is given by:
\begin{equation}
c_n = \dfrac{2^{\frac{n+5}{2}}\Gamma\left(3+\dfrac{n}{2}\right)}{3\sqrt{\pi}}.
\end{equation}
Here $m_H$ is the mass of hydrogen, and $F_\text{He}$ parameterizes the correction to the cross section due to scattering on helium; if there is no DM-helium scattering, then $F_\text{He} = 1 - Y_\text{He }\approx 0.76$, where $Y_\text{He} \approx 0.24$ is the helium mass fraction. 

The temperature evolution for the two populations is governed by the equations \cite{Sigurdson:2004zp,Dvorkin:2013cea}:
\begin{eqnarray}
\dot{T_\chi} &=& - 2 \dfrac{\dot{a}}{a} T_\chi + \dfrac{2 m_\chi}{m_\chi +m_H} R_\chi \left( T_b - T_\chi \right), \nonumber \\
\dot{T_b} &=& - 2 \dfrac{\dot{a}}{a} T_b + 2\dfrac{\mu_b}{m_e} R_\gamma \left( T_\gamma - T_b \right)  \nonumber \\
&& + \dfrac{2 \mu_b}{m_\chi+m_H} \dfrac{\rho_\chi}{\rho_b}R_\chi \left( T_\chi - T_b \right).
\label{tempevol}
\end{eqnarray}
Here $\mu_b$ is the mean molecular weight for the baryons, $\mu_b = m_H (n_H + 4 n_{He})/(n_H+n_{He}+n_e)$.

The evolution of DM temperature, and the time at which the DM temperature deviates from the baryon temperature, depends on $n$. For larger $n$, DM and baryons are coupled with each other early on, but after the scattering rate becomes smaller than the Hubble rate, the DM cools adiabatically with the expansion of the universe. For example, for $n = 0$, the DM-baryon scattering time scale $t_{DB} \equiv (a/R_\chi) \, (m_\chi + m_H)/m_\chi$ sets the decoupling temperature. For this case, we will approximate the DM temperature to be the baryon temperature when $H t_{DB} > 1$, and assume the DM is non-relativistic and so cools adiabatically with $v \propto 1 + z$ after decoupling.

For $n=-4$, this approach fails because the scattering timescale is always longer than the Hubble time at high redshifts (at least within the redshift range we consider). As a calibration point, we can consider the DM temperature at late times if the DM also possesses an annihilation channel which yields the correct thermal relic density; in this case, the decoupling redshift is set by the thermal freezeout condition, $H(z) = (\rho_\chi/m_\chi) \langle \sigma v\rangle$, where $\langle \sigma v \rangle \sim 10^{-26} \text{cm}^3/\text{s}$. For the DM mass range we consider, this condition leads to extremely low DM temperatures at $z < 10^6$, until the DM recouples to the baryons via the scattering interaction. Therefore we will set the DM temperature $T_\chi$ to be 0 K initially, and solve the Boltzmann equations with that initial condition.

Throughout this work, unless specifically noted otherwise, we will neglect scattering on helium and take $F_\text{He} = 0.76$. If helium-DM scattering were to be included, we would need to make the following modifications:
\begin{itemize}
\item $F_\text{He}$ would be given by \cite{Dvorkin:2013cea}:
\begin{align}\,\,\,\,\,\,\,\,\,F_\text{He} & = 1 - Y_\text{He} + Y_\text{He} \times \nonumber \\
&  \frac{\sigma_\text{He}}{\sigma_H}\frac{m_H + m_\chi}{4 m_H + m_\chi}  \left(\frac{T_b m_\chi + T_\chi m_H}{T_b m_\chi + 4 T_\chi m_H} \right)^{(n+1)/2}. 
\label{hecor}
\end{align}
For example, for spin-independent, isospin-independent scattering we expect $\sigma_\text{He}/\sigma_H \approx 2 \mu^2_{\chi \text{He}}/\mu^2_{\chi H}$ \cite{Xu:2018efh}, and so $\sigma_\text{He}/\sigma_H \approx 2$ for $m_\chi \ll m_H$.
\item In the temperature-evolution equations, $R_\chi$ should be replaced by \cite{Dvorkin:2013cea}:
\begin{equation} R_\chi^\prime \equiv R_\chi \left[1 + \frac{3 m_H}{m_\chi + 4 m_H} \left(\frac{1 - Y_\text{He}}{F_\text{He}} -1 \right) \right]\end{equation}
\end{itemize}
Since $F_\text{He} > 1 - Y_\text{He}$, including helium always reduces the effective DM-baryon interaction coefficient in the temperature evolution equations, compared to the coefficient relevant for the evolution of the perturbations; we thus expect models with DM-helium scattering to give rise to a lower temperature distortion for a given modification to the CMB anisotropy spectra. 

When the baryon-photon fluid is tightly coupled at early times, the perturbation equations can be expanded in powers of $\tau_c$ \cite{Sigurdson:2004zp,Xu:2018efh}, yielding:
\begin{eqnarray}
\dot{\theta_b} &=& \dfrac{1}{1+R+\beta R}  \Bigg( -\dfrac{\dot{a}}{a} \theta_b +c_b^2 k^2 \delta_b + R k^2 \left( \dfrac{1}{4}\delta_\gamma-\sigma_\gamma \right)  \nonumber \\
&& + R \dot{S}_{b\gamma} + R \beta \left(\dfrac{\dot{a}}{a} -\dfrac{\dot{\tau_\chi}}{\tau_\chi} \right)\left(\theta_\chi-\theta_b\right)  \nonumber \\
&& + \dfrac{S}{\tau_\chi}\left(\theta_\chi-\theta_b\right) + R \beta \dot{\theta_\chi}\Bigg),\nonumber \\
\dot{\theta_\gamma} &=& -\dfrac{1}{R} \left(\dot{\theta_b} +\dfrac{\dot{a}}{a}\theta_b-c_b^2 k^2 \delta_b^2 \right) + k^2 \left( \dfrac{1}{4} \delta_\gamma - \sigma_\gamma \right) \nonumber \\
&& + \dfrac{S}{R \tau_\chi} \left( \theta_\chi - \theta_b \right),
\end{eqnarray}
where 
\begin{eqnarray}
R &=& \dfrac{4 \rho_\gamma}{3 \rho_b},\nonumber \\
S &=& \dfrac{\rho_\chi}{\rho_b},\nonumber \\
\beta &=& \dfrac{S}{1+R}\dfrac{\tau_\chi}{\tau_b}
\end{eqnarray}
and $\tau_\chi = R_\chi^{-1}$, and $S_{\gamma b} = \theta_\gamma - \theta_b$ describes the standard photon-baryon slip. 

In order to numerically compute the change in the CMB anisotropy spectra, we modify the public code \texttt{CLASS} \cite{lesgourgues2011cosmic} to take into account DM-baryon scattering, via the equations of this section. 

In principle one could also expand the temperature evolution in $\tau_c^{-1}$ in the presence of DM-baryon scattering to track the separate evolution of $T_b$ and $T_\gamma$ in the early universe. We do not modify the standard \texttt{CLASS} treatment which sets the two temperatures equal well before recombination; we expect this approximation to have negligible effect.

\section{Constraints from Anisotropies of the Cosmic Microwave Background}
\label{sec:cmbconstraints}

The CMB anisotropy spectrum can be used to set stringent constraints on interactions between DM and the known particles. For example, DM annihilation (e.g. \cite{Padmanabhan:2005es,Galli:2009zc, Slatyer:2009yq,Slatyer:2015jla,Slatyer:2015kla}) or decay \cite{Chen:2003gz,Slatyer:2016qyl,Poulin:2016anj} to SM particles can inject energy into the photon-baryon fluid between recombination and reionization, heating and ionizing the hydrogen gas and changing the CMB anisotropy spectrum accordingly.

The primary effect of DM-baryon scattering is different; if the DM has non-negligible interactions with the photon-baryon plasma during some epoch, then the pressure from the plasma will reduce the growth of DM overdensities at that redshift. Modes that are within the horizon and growing during this epoch will experience a suppression in their growth relative to the standard calculation, while longer-wavelength modes that enter the horizon later will be less affected. This leads to a suppression of small-scale power in the matter power spectrum and modifications to the CMB temperature and polarization anisotropies, which can be tested against data and constrained.

The effects of DM-baryon scattering on the CMB have been worked out by several authors, for example Refs.~\cite{Sigurdson:2004zp,McDermott:2010pa,Dvorkin:2013cea,Gluscevic:2017ywp,Boddy:2018kfv,Xu:2018efh}.  Most recently, Ref.~\cite{Gluscevic:2017ywp} has computed the limit on velocity-independent scattering from \emph{Planck} 2015 data \cite{Planck:2015xua}  for keV -- TeV DM, with Refs.~\cite{Gluscevic:2017ywp,Boddy:2018kfv,Xu:2018efh} studying velocity-independent and velocity-suppressed ($n > 0$) scattering cross sections for DM masses in the keV-TeV range, and scattering cross sections enhanced at low velocities ($n=-4, -2$) in the MeV-GeV range. Since the effect on the CMB is dominated by small scales, we will investigate the impact of including ACT/SPT data to extend the analysis up to $\ell_\text{max} = 5000$, in particular for the $n=-4$ case where limits from Lyman-alpha have been found to be subdominant to the CMB bounds \cite{Xu:2018efh} . 

One assumption we made when deriving the CMB perturbations in Sec.~\ref{sec:formalism} is that the DM-baryon relative bulk velocity $V_\text{rms}$ is small compared with the thermal velocity, so that the scattering coefficient $R_\chi$ is approximately independent of the velocity divergences $\theta_b$ and $\theta_\chi$. The rms value of this relative velocity is given by \cite{2010PhRvD..82h3520T}:
\begin{equation}
V_\text{rms}^2=\int \dfrac{d k}{k} \Delta_\zeta \left(\dfrac{\theta_b-\theta_\chi}{k}\right)^2.
\end{equation} 
where $ \Delta_\zeta$ is the primordial curvature perturbation $\sim 2.4 \times 10^{-9}$ per log $k$. Thus a full treatment of this effect would require accounting for the presence in $R_\chi$ of the $\theta$ functions for different $k$-modes, leading to coupled evolution between modes with different $k$.

In order to avoid the necessity of treating coupled $k$-modes in the low-redshift regime where $V_\text{rms}$ is important, we follow Ref.~\cite{Dvorkin:2013cea}, which made the replacement:
\begin{equation}
R_\chi \rightarrow \dfrac{a c_n \rho_b \sigma_0 }{m_\chi+m_H} \left( \dfrac{T_b}{m_H}+\dfrac{T_\chi}{m_\chi} + \dfrac{V_\text{rms}^2
}{3}\right)^{\frac{n+1}{2}}F_\text{He},\label{eq:rchi_with_vrms}
\end{equation}
where $V_\text{rms}$ is estimated as:
\begin{align}V^2_\text{rms} \approx \begin{cases} 10^{-8} & z > 10^3 \\ 10^{-8} \left(\frac{1+z}{10^3}\right)^2 & z \le 10^3.\end{cases} \end{align}
Ref.~\cite{Dvorkin:2013cea}, argues that this ``mean-field'' approach is valid for $z > 10^4$, so can be used for scattering models where the signal is dominated by $z > 10^4$. For models that are strongly enhanced at low velocities, $n < -2$, the signal may peak at low redshifts, but in that case this treatment should be conservative, in the sense that a more detailed treatment would be likely to yield even stronger constraints. We will later discuss alternative treatments of the DM-baryon relative velocity in the context of late-time cooling of the baryons through scattering.

\subsection{Results with and without ACT/SPT data}

\begin{figure}[h]
     \includegraphics[width=9.cm]{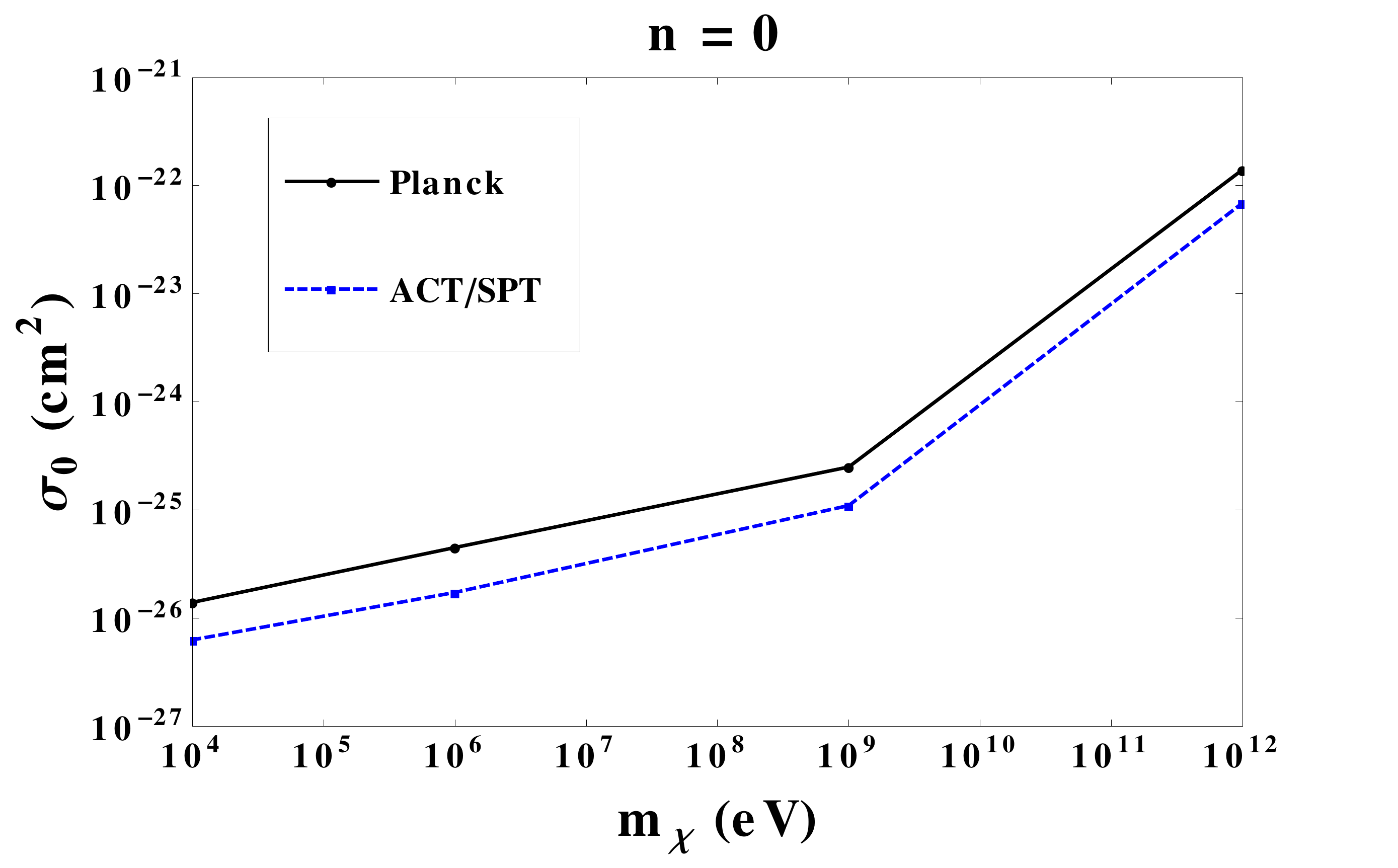}
     \includegraphics[width=9.cm]{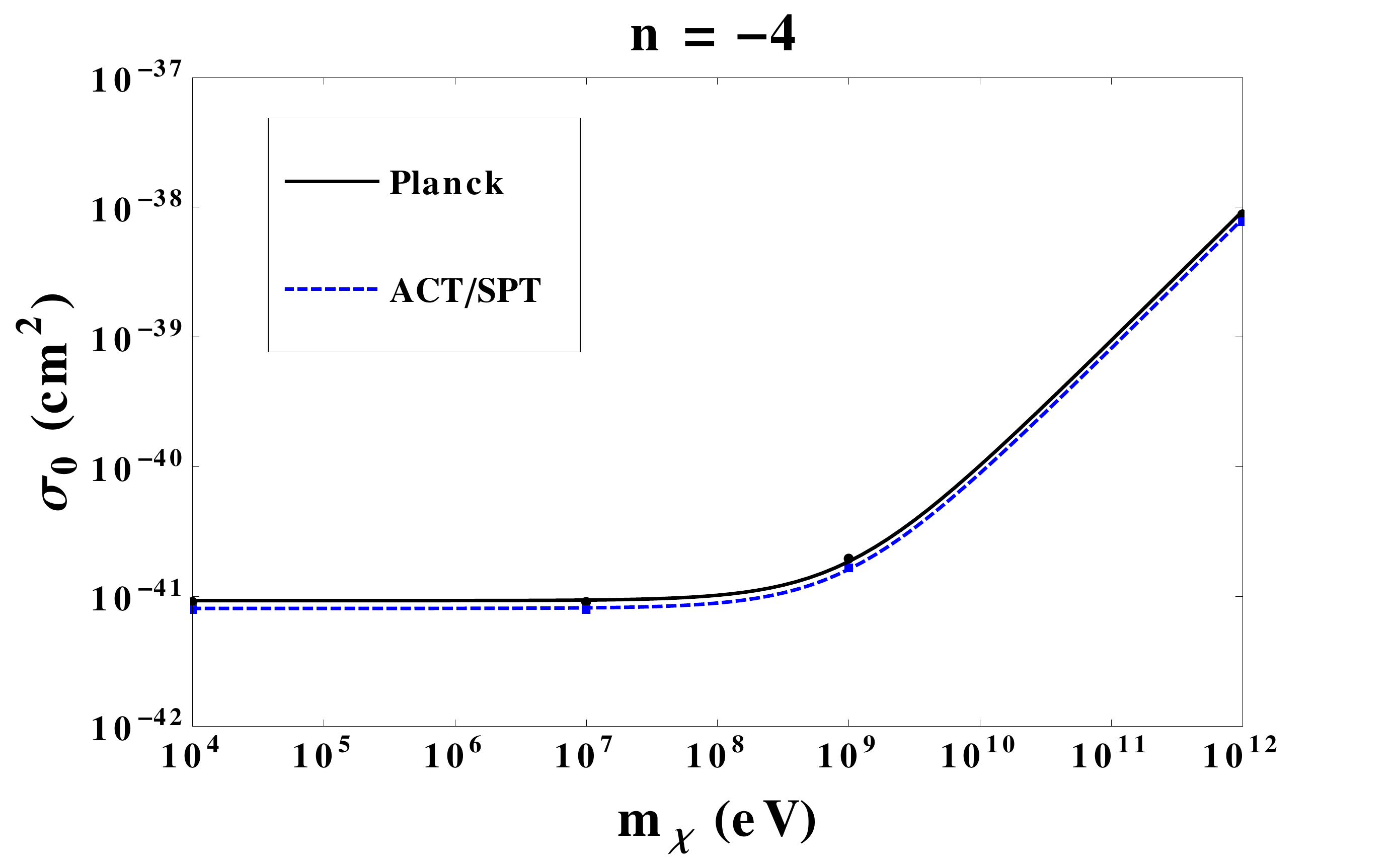}
  \caption{
   95\% confidence upper limits on the DM-baryon scattering cross section $\sigma = \sigma_0 v^n$, computed using \texttt{MontePython}, for (\emph{upper panel}) $n = 0$ and (\emph{lower panel}) $n=-4$. For \emph{black, solid} lines labeled ``Planck'' the \emph{Planck} 2015 TTTEEE likelihood is employed; for \emph{blue, dashed} lines labeled ``ACT/SPT'' the 2013 ACT and SPT likelihoods are added. Dots indicate MCMC results. In the \emph{upper panel}, these dots are joined by straight lines; in the \emph{lower panel}, the line follows the $\sigma_0 \propto 1 + m_\chi/m_H$ scaling suggested in Ref.~\cite{Xu:2018efh}.}
\label{CMBlimit}
\end{figure}

The limits on scattering from \emph{Planck} 2015 data for $n = 0$ and $n = -4$ are shown in Fig.~\ref{CMBlimit}. These limits were computed using the MCMC code \texttt{MontePython} \cite{audren7183conservative} and the full \emph{Planck} 2015 likelihood (TT + TE + EE, low-$\ell$ and high-$\ell$, using lensed $C_\ell$'s and the lensing likelihood) \cite{Aghanim:2015xee}, floating all six standard cosmological parameters in addition to the scattering cross section. We do not consider here constraints from Lyman-$\alpha$ data, but previous studies have found that these constraints are important for the $n=0$ case, but subdominant for $n=-4$ scattering \cite{Xu:2018efh}. In the $n=-4$ case we compare our MCMC results to a simple mass-scaling relationship suggested by \cite{Xu:2018efh}, with the limiting cross-section $\sigma_0 \propto 1 + m_\chi/m_H$, and find good agreement.

To include ACT/SPT high-$\ell$ data, covering the range from $\ell \sim 1500-5000$, we employ the 2013 ACT/SPT likelihoods \cite{Das:2013zf,Dunkley:2013vu,Calabrese:2013jyk} as implemented in \texttt{MontePython}, in addition to the \emph{Planck} TTTEEE likelihood.

\begin{figure}[h]
     \includegraphics[width=9.cm]{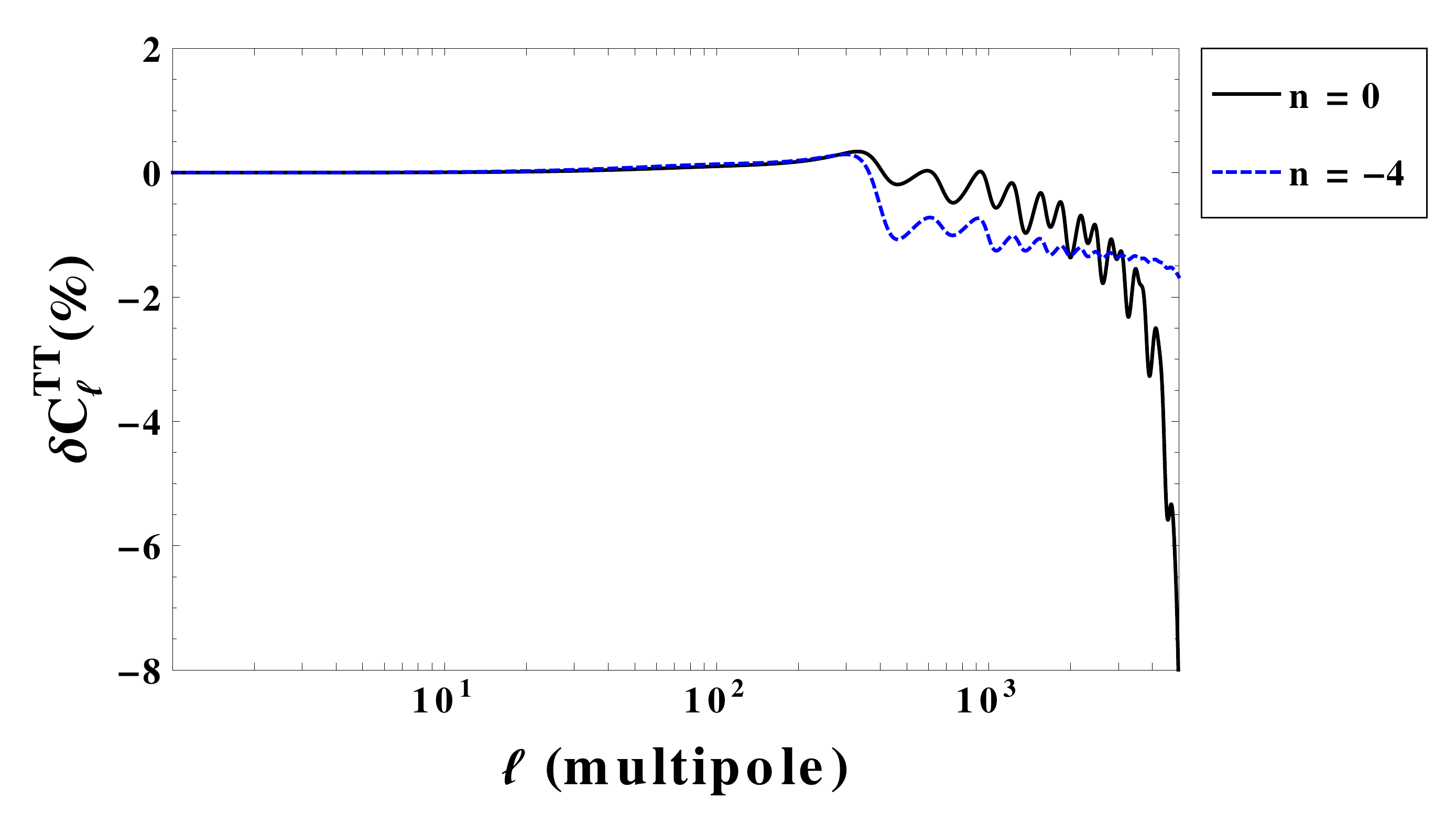}
  \caption{The change in the temperature power spectrum up to $\ell = 5000$, assuming DM mass to be 0.1 GeV, with a cross section given by the ``\emph{Planck}'' upper limit in Fig.~\ref{CMBlimit}. }
\label{planckcl}
\end{figure}

We find that adding these data improves the limit by about $10\%$ for the $n=-4$ case (and about a factor of 2 for $n=0$, albeit the Lyman-alpha constraints \cite{Xu:2018efh} are still stronger in this case), in a largely mass-independent way. In Fig.~\ref{planckcl} we plot the modifications to the CMB temperature anisotropies (holding cosmological parameters constant at their $\Lambda$CDM best-fit values for this demonstration; they are floated in the likelihood scan) for the maximum cross section allowed by \emph{Planck}. 

\subsection{Testing linearity}
\label{sec:linearity}

In the case of exotic energy injections, the effect on the CMB anisotropies is approximately linear in the energy injection \cite{Finkbeiner:2011dx}. Therefore, the effect of a general energy deposition history can be described in terms of a linear combination of basis energy deposition histories, and we can use Fisher forecasting and principal component methods to accurately estimate the effects of arbitrary energy injection histories. In this section, we demonstrate that a similar statement can be made for the effects of scattering (in particular $n=-4$ scattering) on the CMB anisotropies.

\begin{figure}[h]
     \includegraphics[width=9cm]{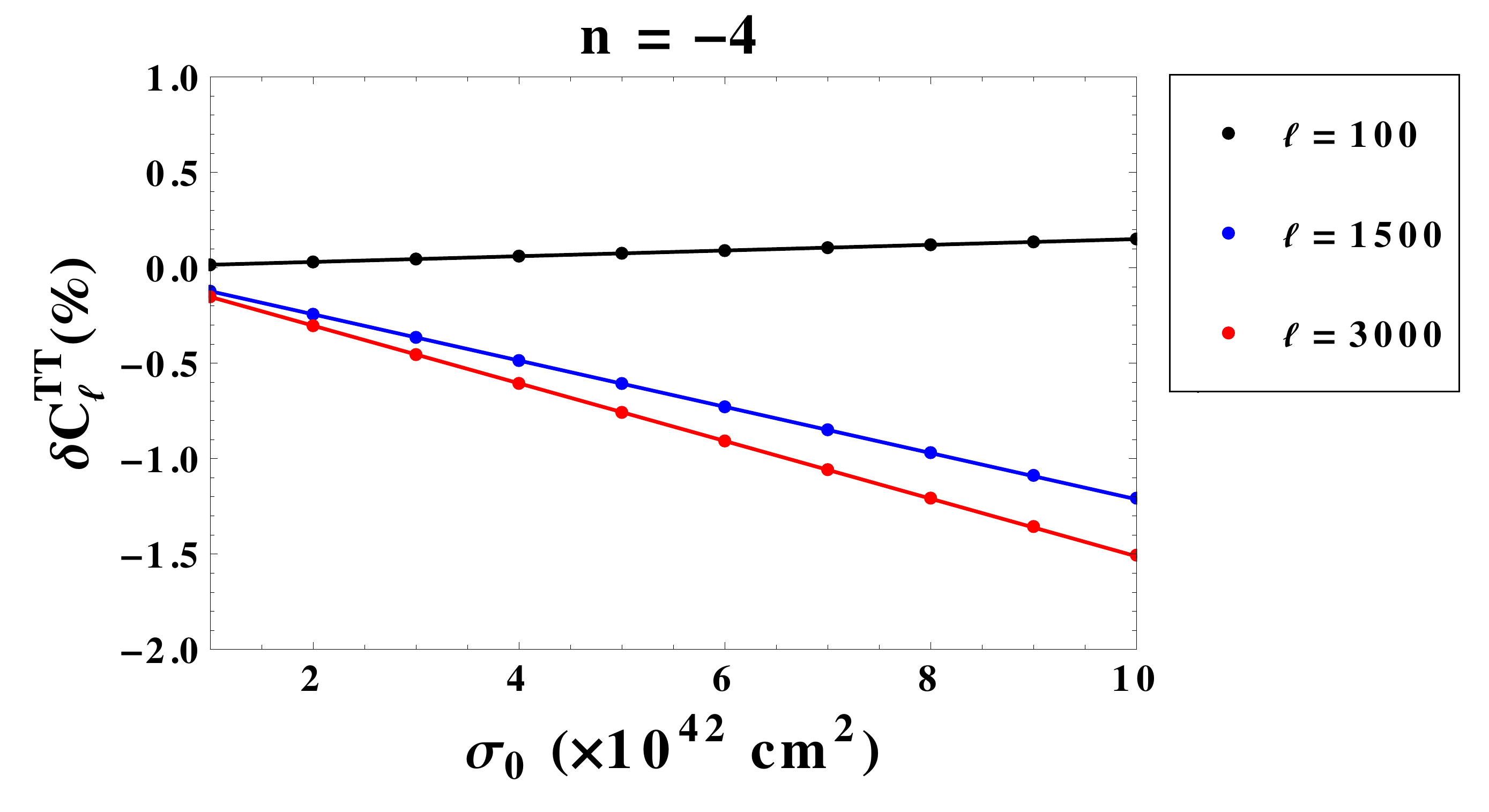}
  \caption{
   Change of the CMB temperature power spectrum at $\ell = 100, 1500, 3000$ as a function of the DM-baryon scattering cross section, for a DM mass of 0.1 GeV; cosmological parameters are held fixed at the $\Lambda$CDM values.}
\label{linearity}
\end{figure}

\begin{figure}[h]
     \includegraphics[width=9cm]{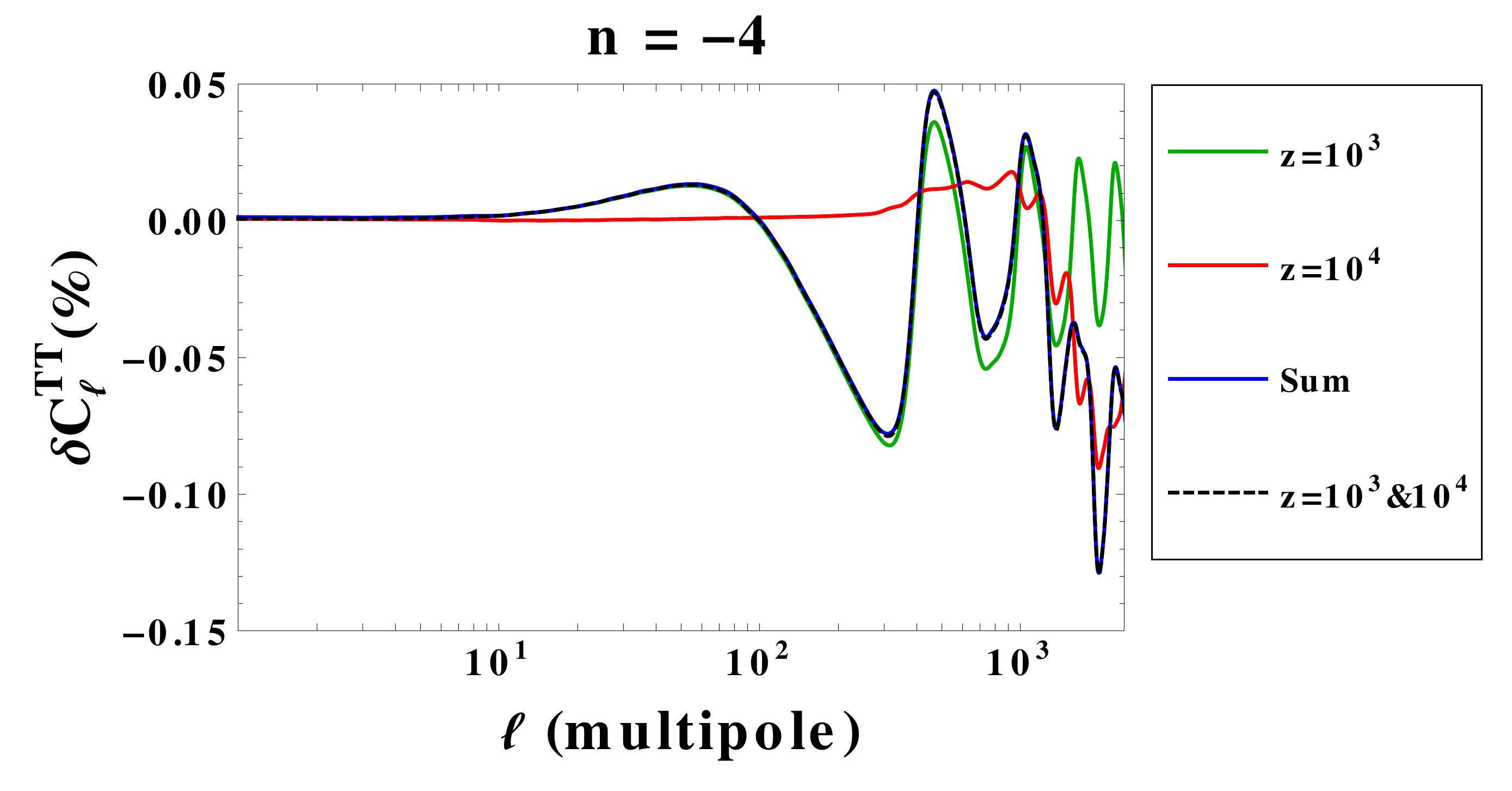}
  \caption{
   The effect on the CMB temperature power spectrum of turning on DM-baryon scattering in a narrow redshift range about $10^3$ (\emph{green} line), $10^4$ (\emph{red} line), and at both redshifts (\emph{black dashed} line). The cross section is set to $\sigma_0  = 2\times 10^{-41} \left( \text{cm}^2\right)$ for $z = 10^3$ and $\sigma_0  = 10^{-41} \left( \text{cm}^2\right)$ for $z = 10^4$, and in both cases $\sigma = \sigma_0 v^{-4}$ scaling is assumed. The sum of the red and green lines is shown as the \emph{blue line}; to the degree that the problem is linear, the blue line should overlap the black dashed line.}
\label{linearity2}
\end{figure}

One simple test of linearity is the degree to which the change in the anisotropy power spectrum, $\delta C_{\ell}$, is linear with respect to the DM-baryon scattering cross section. At sufficiently large cross sections linearity will necessarily break down, but we find that it is a good approximation for cross sections that are not excluded by the CMB constraints discussed in the previous section. The $\delta C_{\ell}$ changes as a function of cross section are shown in Fig.~\ref{linearity}, for several sample multipoles. The degree of non-linearity is smaller than $10\%$ up to a cross section of $\sigma_0 \approx 10^{-40} \, \text{cm}^2$, for low-mass DM (below 1 GeV). Another test is whether the effect of scattering at two different redshifts is the same as the sum of the effects of scattering at the two redshifts individually; we test this in Fig.~\ref{linearity2}. We find good agreement, indicating that for these cross sections and for $n=-4$ velocity scaling, it is reasonable to consider the effect of scattering over a wide redshift range as a linear combination of the individual effects of scattering over smaller redshift ranges.

We note that linearity would be expected to break down if the $T_\chi/m_\chi$ term appearing in the scattering coefficient $R_\chi$ (Eq.~\ref{eq:rchi_with_vrms}) becomes non-negligible compared to the $T_b/m_H$ and $V_\text{rms}^2/3$ contributions, in the case that the DM temperature $T_\chi$ is itself a function of the scattering cross section. For the case with $n=-4$, this is not a concern for cold dark matter within our formalism and the range of masses we consider; the DM-baryon scattering is never coupled in the early universe, the initial DM temperature is assumed to be very small, and consequently the $T_\chi/m_\chi$ term is always subdominant (as also discussed in \cite{Xu:2018efh}). In contrast, for $n \ge -2$ the DM was initially coupled to the baryons through scattering and subsequently decoupled, and for sufficiently light DM, it is possible for the $T_\chi/m_\chi$ term to dominate during the redshifts relevant to the CMB. Since the DM temperature is always less than or equal to the baryon temperature within our framework, this behavior occurs only for DM masses below 1 GeV.

As an example, for 10 MeV DM with a scattering cross section of $10^{-25}$ cm$^2$, comparable to the \emph{Planck} limit for $n=0$ scattering shown in Fig.~\ref{CMBlimit}, the decoupling redshift is $z\sim 5\times 10^5$. We will argue in the next section that the CMB constraints are dominated by scattering occurring in the redshift range $z\sim10^{3}-10^4$. After decoupling, the DM cools faster than the CMB by a factor of $(1+z)$, so at $z\sim 10^3-10^4$, the DM is 50-500$\times$ cooler than the baryons. Since the DM is roughly $100\times$ lighter than $m_H$, over this epoch, the $T_\chi/m_\chi$ term is comparable to the $T_b/m_H$ term (being larger initially and smaller at late times). Both terms are also comparable to the $V_\text{rms}^2/3$ term. 

For lighter DM, the effect of the $T_\chi/m_\chi$ term will be more pronounced. The redshift factor between decoupling and the CMB epoch, and hence the ratio of the CMB temperature to the DM temperature over the time of interest, scales as $m_\chi^{-1/3}$ for $m_\chi \ll m_H$; thus the ratio $(T_\chi/m_\chi)/(T_b/m_H)$ scales as $m_\chi^{-2/3}$. For heavier DM, the effect will be small.

This effect is responsible for the change in the shape of the constraints on $\sigma$ as a function of $m_\chi$ between the $n=0$ and $n=-4$ cases visible in Fig.~\ref{CMBlimit}. We observe that for $n=0$, the constraints continue to strengthen as the DM mass drops further below 1 GeV, whereas for $n=-4$ they asymptote to a constant value. The reason is that in the $n=0$ case, when the DM temperature becomes important it increases the relative velocity between DM and baryons, and this corresponds to an increased scattering rate and a correspondingly lower allowed cross-section. From the argument above, we anticipate that the deviation from the naive scaling should begin to be appreciable around $m_\chi \sim 10$ MeV; we will confirm and quantify this shortly.

\subsection{Characterizing the redshift of interest for the CMB constraints}
\label{sec:redshift}

As discussed previously, the mechanism for distorting the CMB through baryon-DM scattering is that transfer of energy from the baryons to the DM modifies the growth of matter perturbations, which in turn is imprinted onto the CMB. After recombination, the decoupling of the photon and baryon fluids reduces or eliminates the imprint of subsequent scattering on the CMB. To see explicitly which redshifts dominate the CMB signal, we can consider turning on DM-baryon scattering for short periods prior to recombination, and studying the impact on the CMB; given the linearity results of the previous section, the final signal can be estimated as the sum of these localized-in-redshift scattering histories. This is intended only as a means to explore the varying effects of scattering at different redshifts -- most physical models for DM-baryon scattering will have scattering over a wide range of redshifts -- but it can be quite a good approximation to models where the scattering rate rises steeply at low redshifts, before the signal is cut off by recombination. In principle, velocity- or temperature-dependent resonance effects could also enhance scattering at particular redshifts.

As noted previously, this approach cannot always be applied to light sub-GeV DM where the DM has previously been more strongly coupled to the baryons (as is the case for a $n \ge -2$ power-law dependence on velocity); in this case the $T_\chi/m_\chi$ term in $R_\chi$ can become important if the DM is sufficiently light, and the thermal history of the dark matter -- which in general will have been affected by previous epochs of scattering -- must be specified in order to compute the effect on the CMB perturbations.

We consider a scattering history $\sigma(z)$ starting with $\sigma = \sigma_0 v^{-n}$, but then modulate this history by a redshift-dependent Gaussian function, $G_i(z) \propto e^{-(z-z_i)^2/(2 \Delta z_i^2)}$, peaked at a central redshift $z_i$ and with a width parameter $\Delta z_i$. We choose the $z_i$ values to be linearly spaced between $z=10^2$ and $z=5\times 10^4$. We choose $\Delta z_i$ to be the spacing between adjacent $z_i$, and normalize the Gaussians such that $\int dz G_i\left(z\right) = \Delta z_i$ (i.e. they have the same normalization as a step function that is 1 in the range $z_i \pm (\Delta z_i)/2$). Thus summing together all these modulated scattering histories approximately recovers the original scattering history.

We calculate the perturbation to the CMB anisotropy spectra for each choice of $z_i$ as discussed earlier. Note that for the $n=0$ case, we must make a decision as to what initial conditions to impose on the DM temperature for each modulated scattering history. Two simple options are (1) set the initial DM temperature $T_\chi=0$ (as in the $n=-4$ case), (2) assume that there was an earlier period of $n=0$ scattering that coupled the DM temperature to the baryon temperature until the scattering timescale became comparable to the Hubble time. 

The latter prescription ensures that the DM temperature in each of the modulated scattering histories is similar to the DM temperature at the same redshift in the combined history. This is appropriate when decomposing a full $n=0$ history into individual redshift slices, but it means that a general redshift-dependent scattering history (which could lead to a very different thermal history) cannot generally be decomposed into a linear combination of these modulated histories. Likewise, for $n=0$ and light DM, where the problem is not expected to be linear, the relative effects of scattering at different redshifts will in general depend on the assumed cross section for the baseline scattering history (since this sets the time of decoupling), if the second approach is taken. The first prescription tends to ensure a very low DM temperature at all times, preserving linearity of the problem, but it may not be self-consistent if the overall scattering history is sufficient to appreciably heat the dark matter.

These prescriptions give equivalent results, and a general redshift-dependent scattering history can be built up by taking linear combinations of the modulated scattering histories, if the DM temperature remains low enough that the scattering rate is approximately independent of the DM temperature. As discussed previously, we find this is generically the case for 0.1 GeV and heavier DM, or for cases where the initial DM temperature is very low and the DM has not been strongly coupled to baryons early in the universe.

Thus for this analysis, for the $n=-4$ case we test three DM masses, 10 keV, 100 MeV and 1 TeV, to demonstrate the level of variation in redshift dependence with DM mass; we will subsequently test the effects of varying the DM mass for fixed redshift dependence. For the $n=0$ case, we restrict our attention to the 100 MeV and 1 TeV cases, where the prescriptions above give equivalent results.

In the case of $n=0$ scattering with 10 keV (or similarly low-mass) DM, the problem becomes much more complicated, as the redshift-dependence of the signal is now a function of the decoupling temperature and hence of the scattering cross section. Since the $n=0$ case is not the main focus of this paper, we leave further study of this case to future work.

To estimate the significance of such a perturbation as a function of $z_i$, we use a Fisher-matrix-based approach, following standard methodology (see e.g. Ref.~\cite{Finkbeiner:2011dx} for details beyond those presented here).

Let $\alpha_i$ be a coefficient modulating $G_i(z)$, and let us write $\frac{\partial C_\ell}{\partial \alpha_i} = \left\{ \frac{\partial C_\ell^{TT}}{\partial \alpha_i}, \frac{\partial C_\ell^{EE}}{\partial \alpha_i}, \frac{\partial C_\ell^{TE}}{\partial \alpha_i} \right\}$. We use the covariance matrix for the $C_\ell$'s (e.g. \cite{jungman96b,tegmark97,Verde:2009tu}):
\begin{align} \Sigma_\ell & = \frac{2}{2l + 1} \times \nonumber \\
& \begin{pmatrix} (C_\ell^{TT})^2 & (C_\ell^{TE})^2 & C_\ell^{TT} C_\ell^{TE} \\ (C_\ell^{TE})^2 & (C_\ell^{EE})^2 & C_\ell^{EE} C_\ell^{TE} \\ C_\ell^{TT} C_\ell^{TE} & C_\ell^{EE} C_\ell^{TE}  & \left[ (C_\ell^{TE})^2 + C_\ell^{TT} C_\ell^{EE} \right] \end{pmatrix}.\end{align} 
To account for noise, in these expressions we replace $C_\ell^{TT, EE} \rightarrow C_\ell^{TT, EE} + N_\ell^{TT, EE}$, where $N_\ell^{TT, EE} = (\Delta T \times \text{FWHM})^2 e^{l(l+1)\theta^2}$, $\theta$ and $\text{FWHM}$ describe the beam width ($\text{FWHM}= \theta \sqrt{8 \ln 2}$), and $\Delta T$ describes the instrument sensitivity. To account for fractional sky coverage, we also divide $\Sigma_\ell$ by $f_\text{sky}$. To describe a \emph{Planck}-like mission, we take $\text{FWHM} = 7.1$ arcmin, $\Delta T/T = 2.2\times 10^{-6}$ for temperature and $4.2\times 10^{-6}$ for polarization, and $f_\text{sky} = 0.65$.

The (pre-marginalization) Fisher matrix is then obtained by $(F_e)_{ij} =\sum_\ell \left(\frac{\partial C_\ell}{\partial \alpha_i} \right)^T \Sigma_\ell^{-1} \left(\frac{\partial C_\ell}{\partial \alpha_j} \right)$. Marginalization over the cosmological parameters is performed as in Ref.~\cite{Finkbeiner:2011dx}, by computing the derivatives (about the best-fit CDM point) of the $C_\ell$'s with respect to variations in the cosmological parameters, building and inverting an expanded Fisher matrix that includes the cosmological parameters, and extracting the marginalized Fisher matrix for the $\alpha_i$ parameters.

\begin{figure}[h]
     \includegraphics[width=9cm]{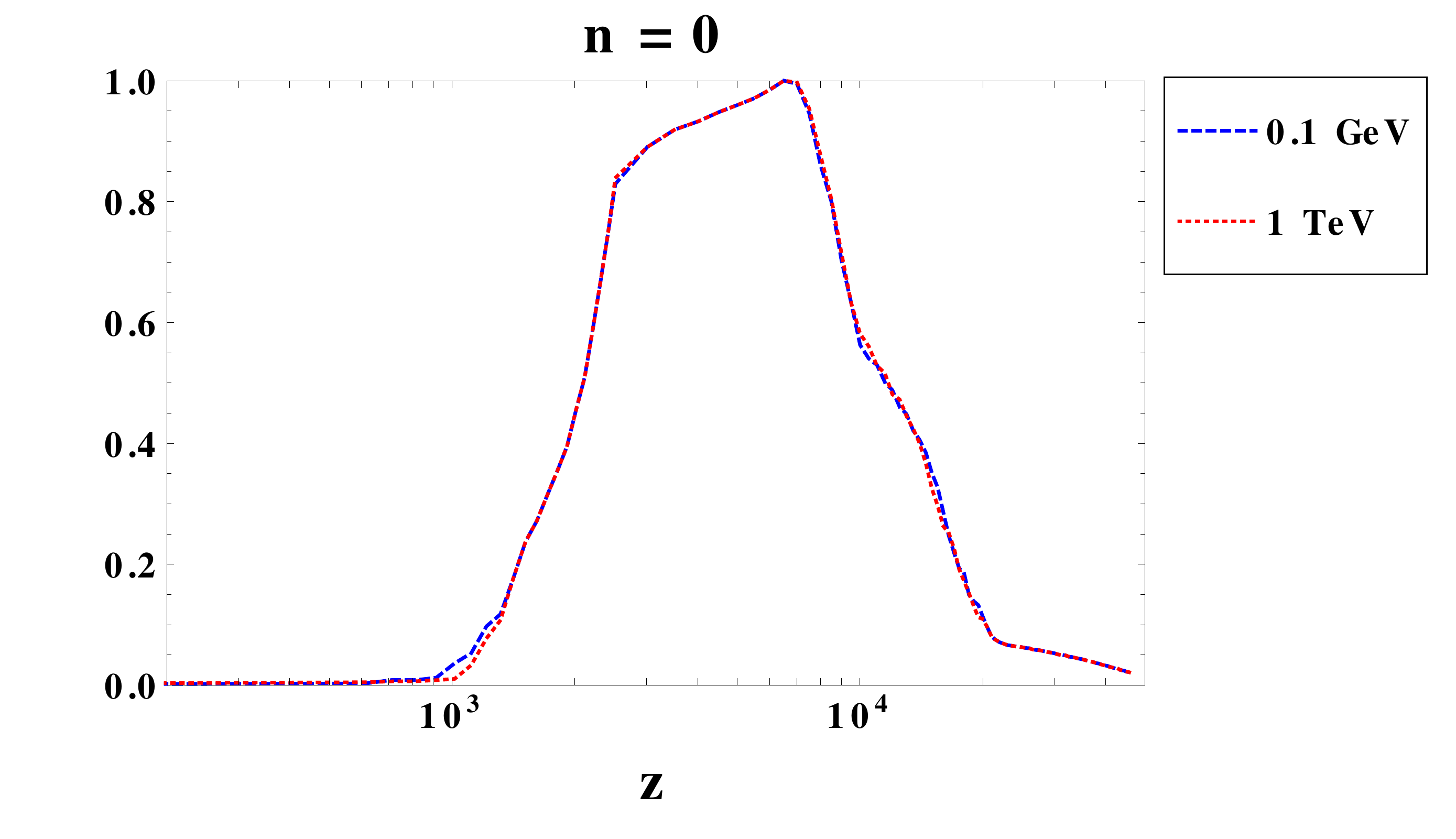}
     \includegraphics[width=9cm]{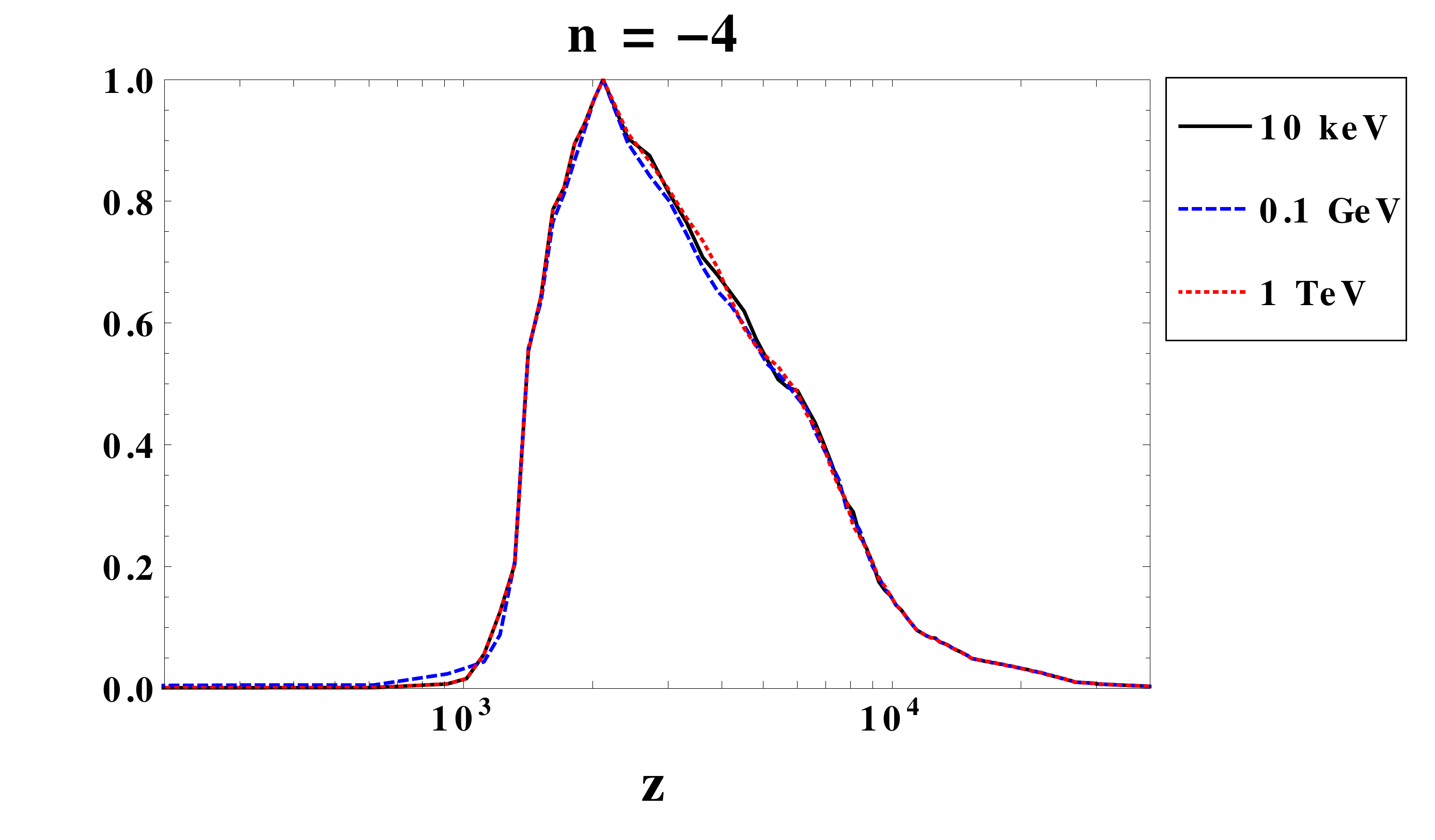}
  \caption{
  Relative significance of isolated scattering at different redshifts, estimated using the Fisher-matrix analysis detailed in the text, for DM masses 10 keV (\emph{solid black line}, lower panel only),  0.1 GeV (\emph{dashed blue line}) and 1 TeV (\emph{dotted red line}), for $n = 0$ (\emph{upper panel})  and $n = -4$ (\emph{lower panel}).}
\label{zscanmasstest}
\end{figure}

Armed with this marginalized Fisher matrix, we can first estimate the relative significance of independent scattering at different redshifts by plotting the $F_{ii}$ terms as a function of $z_i$. The results are shown in Fig.~\ref{zscanmasstest}. We find that for $n=0$ scattering, i.e. where the scattering cross-section $\sigma$ is constant with respect to velocity, the significance is broadly peaked around redshifts of several thousand; for $n=-4$ scattering, since the relative significance of perturbations at lower redshifts is enhanced by the velocity dependence, the peak of significance is sharper and at somewhat lower redshift, at $z\sim 2\times10^3$, and very little signal is produced prior to $z\sim 10^4$. These general statements hold for both light and heavy DM masses, provided that linearity holds.

We note that for $z > 10^3$, the significance curve for $n=-4$ scattering is reasonably well approximated by the $n=0$ curve (for high DM masses where linearity is expected to hold) multiplied by $v^{-4}$, where $v \equiv (T_\chi/m_\chi + T_b/m_H + V_\text{rms}^2/3)^{1/2}$, as one would expect from linearity considerations.

That the most important redshift range for CMB constraints is $z\sim 10^3-$few $\times10^4$ is not surprising; the modes corresponding to the $\ell$ range best-measured by the CMB, up to $\ell$'s of a few thousand, cross the horizon during this epoch. Modifications to the perturbations at earlier times will primarily affect smaller scales, which may be probed by measurements of the matter power spectrum, but not by the CMB.

We note that this justifies the extension of our constraints down to masses below the MeV scale; Ref.~\cite{Xu:2018efh} argued that for sub-MeV masses, relativistic dynamics would need to be included at high redshift $z\sim 10^9$. However, since the signal appears to be almost entirely set by redshifts below $z\sim 2 \times 10^4$ (corresponding to a CMB temperature $\sim$ 10 eV), there should be little error in the constraints provided the DM is cold and non-relativistic during this epoch. This should hold true for keV and heavier DM, since the DM temperature never exceeds the CMB temperature as a result of DM-baryon scattering.

We can also diagonalize this marginalized Fisher matrix to obtain principal components. This analysis decomposes the space of perturbations to the CMB anisotropy spectrum due to scattering at different redshifts into a set of orthogonal basis vectors; to the degree that the problem is approximately linear, the impact of an arbitrary redshift-dependent scattering history on the CMB can be obtained by decomposing that scattering history $\sigma(z)$ into a linear combination of the principal components.

 \begin{figure}[h]
     \includegraphics[width=9cm]{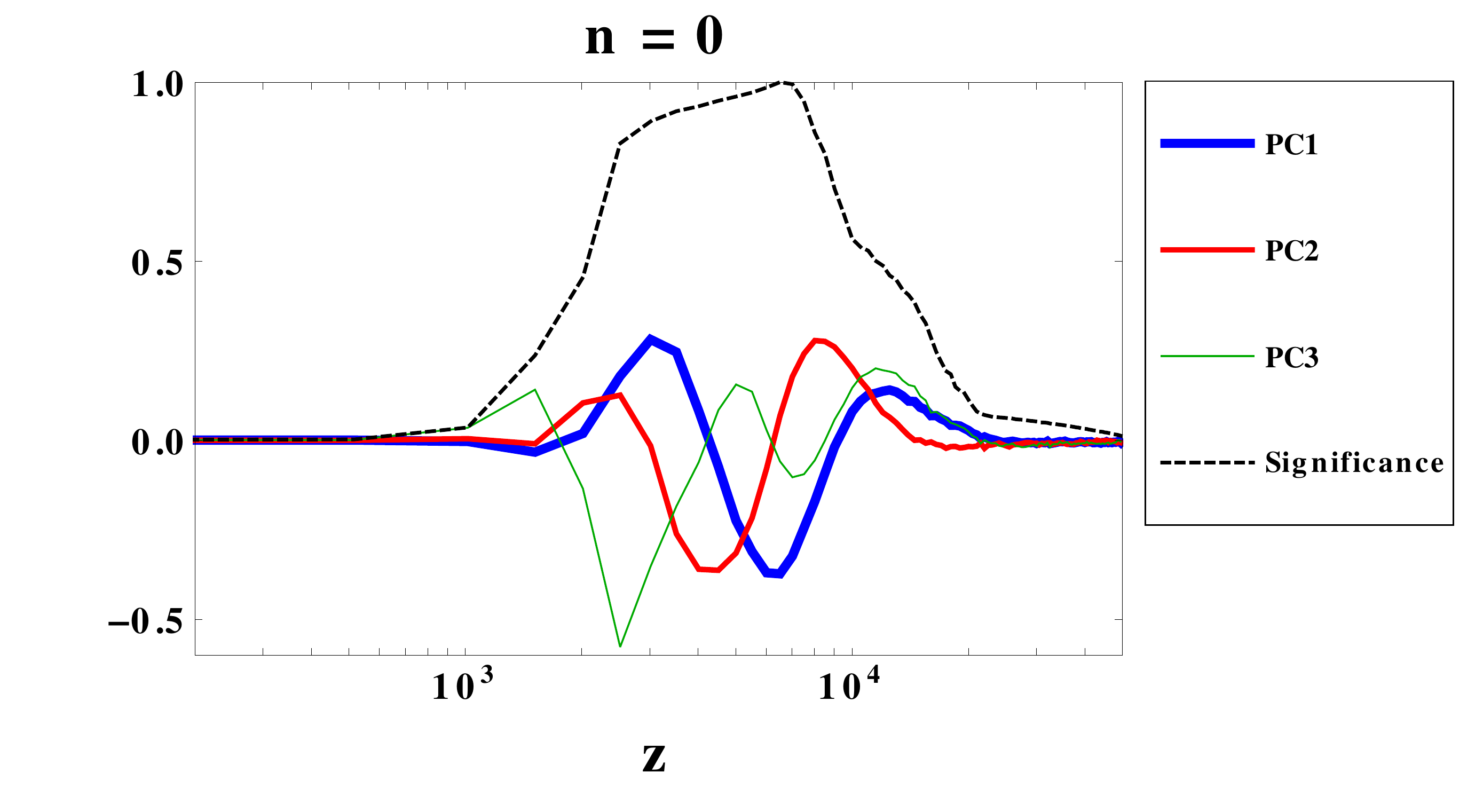}
     \includegraphics[width=9cm]{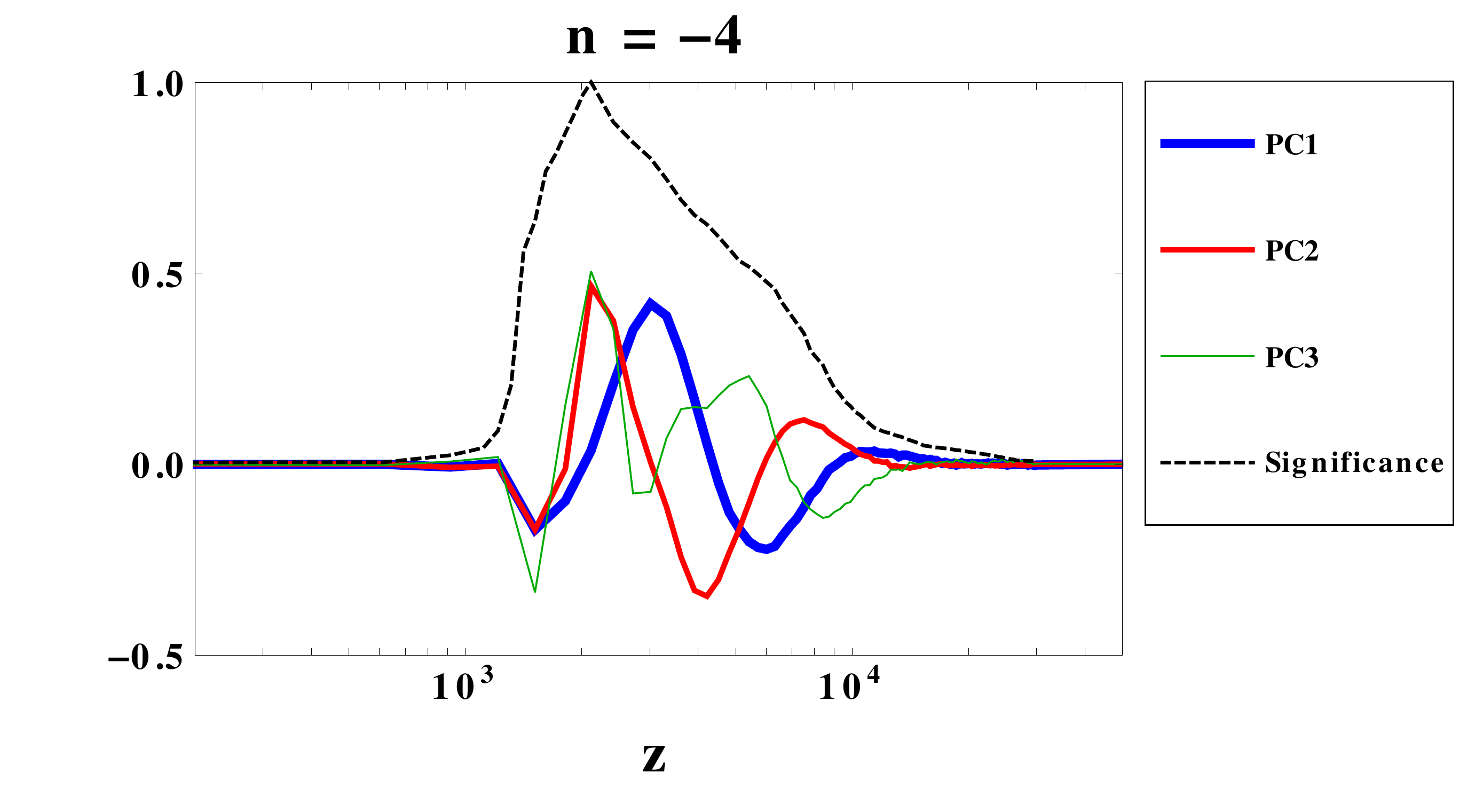}
  \caption{
   Principal components (first = \emph{\{thick, blue\}}, second = \emph{\{medium, red\}}, third = \emph{\{thin, green\}}) for 0.1 GeV DM with redshift-dependent scattering, for $n = 0$  and $n = -4$, overlaid with the significance curves of Fig.~\ref{zscanmasstest}.}
\label{zscan}
\end{figure} 

We display the principal components for 0.1 GeV DM in Fig.~\ref{zscan}; we expect the results to be near-identical for heavier DM masses (or for all DM masses in a $n=-4$-like case where the DM temperature remains very small). Consistent with our significance analysis, we find that the first three principal components have support primarily in the redshift range $10^3 < z < 10^4$. However, in contrast to the cases of DM annihilation \cite{Slatyer:2015jla} or decay \cite{Slatyer:2016qyl}, where the first principal component dominates and the space of perturbations to the CMB is approximately one-dimensional, in both these cases the first principal component only accounts for about $40\%$ of the variance. Thus the space of perturbations to the CMB from scattering is genuinely multidimensional. The reason is that scattering at a given redshift modifies the perturbations at  scales that are inside the horizon at that redshift, so scattering at two different redshifts modifies the power spectrum over two different ranges of $\ell$, leading to different characteristic patterns of modifications of the $C_\ell$'s.

The first four parameters account for roughly $90-95\%$ of the variance (the first two account for roughly $75\%$ of the total in the $n=0$ case and $65\%$ of the total in the $n=-4$ case); thus if $\mathcal{O}(10\%)$ uncertainties are acceptable, the space of scattering histories may approximately be described in terms of four parameters.\footnote{It may still be possible to characterize this space with a smaller number of parameters; we defer this investigation to future work.} To facilitate studies of modified scattering histories, we provide the first four principal components for both cases ($n=0$ and $n=-4$) and a summary of the method for estimating constraints on arbitrary scattering histories in Appendix~\ref{app:fourpcs}.

\subsection{Characterizing mass dependence in the CMB constraints}
\label{sec:mass}

Instead of fixing the DM mass and varying the redshift at which scattering is turned on, we can hold the redshift-dependence of the scattering cross section constant and perform a principal component analysis to study the effects of varying the DM mass between 1 keV and 1 TeV. In this case we find that for $n=-4$, the first principal component describes over 99.9\% of the variance, whereas in the $n=0$ case the ratio is 97\%. Thus (within the limitations of this linear analysis) varying the DM mass is predicted to have very little effect on the shape of the perturbations to the CMB anisotropy spectra, and the main effect is simply to change the overall normalization.

\begin{figure}[h]
 	\includegraphics[width=9cm]{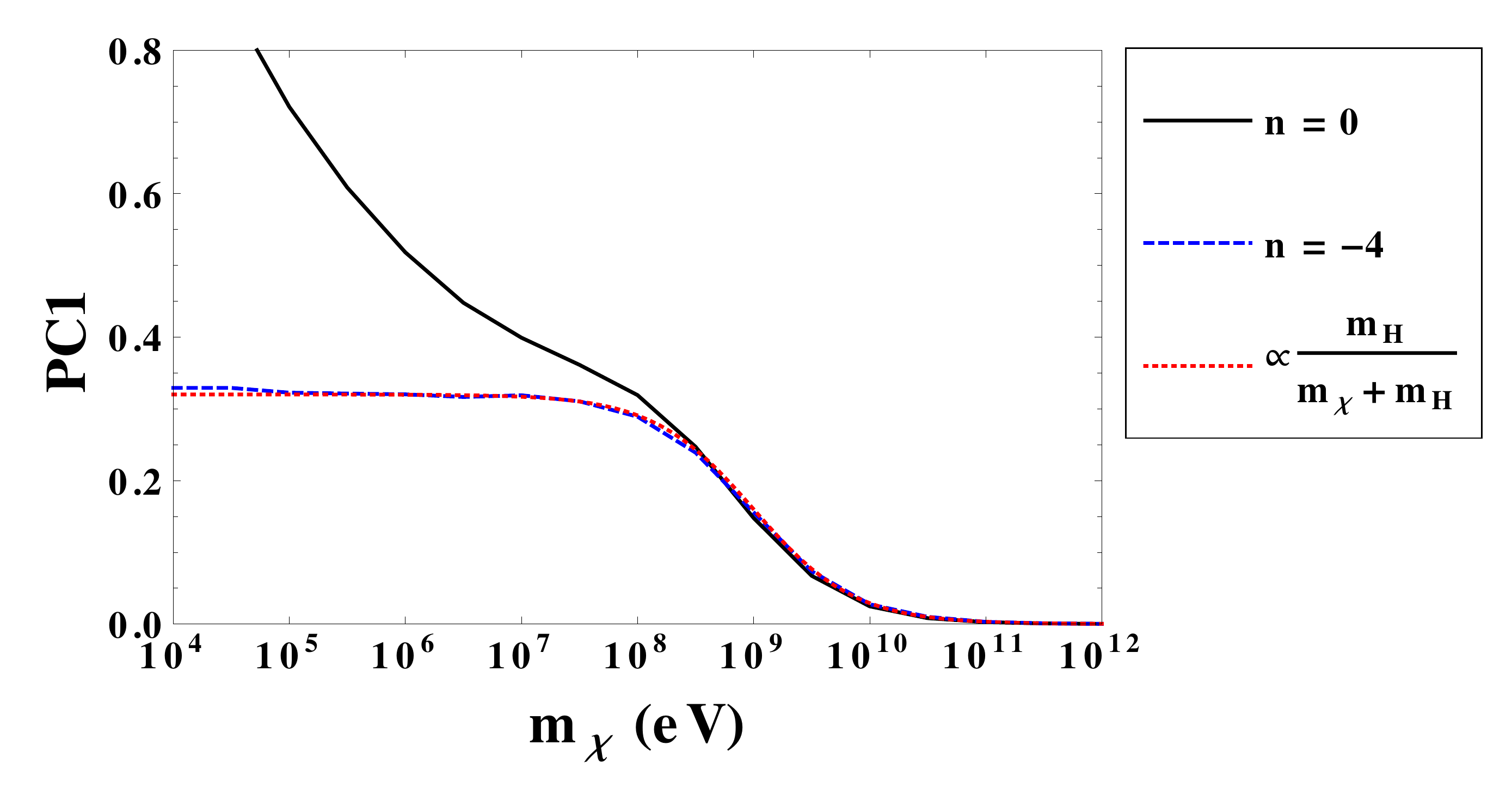}
  \caption{
   First principal component (arbitrary normalization) as a function of the DM mass, for $n=0$ scattering (\emph{solid black line}) and $n=-4$ scattering (\emph{dashed blue line}). We also overplot the simple mass scaling suggested in Ref.~\cite{Xu:2018efh}.}
\label{masspca}
\end{figure}

We plot the first principal component as a function of DM mass in Fig.~\ref{masspca}. In agreement with Ref.~\cite{Xu:2018efh}, we find that for $n=-4$ the shape of this curve is well described by $\mu/m_\chi = m_H/(m_\chi + m_H)$, i.e. to a good approximation the signal scales as the momentum transfer per scattering (proportional to $\mu$) multiplied by the DM number density (proportional to $1/m_\chi$, since the mass density is known but the number density is not). For $m_\chi \ll m_H$, the signal is nearly independent of the DM mass (to the percent level).

For the $n=0$ case, the situation is different because of the dependence on the DM scattering rate on the DM temperature. For masses below about 100 MeV, making the DM lighter increases the DM-baryon relative velocity and hence the scattering rate, causing a divergence from the simple scaling above that becomes increasingly pronounced at lower DM masses. When this effect becomes large, the linearity of the problem also breaks down, invalidating the PCA approach.

\begin{figure}[h]
 	\includegraphics[width=9cm]{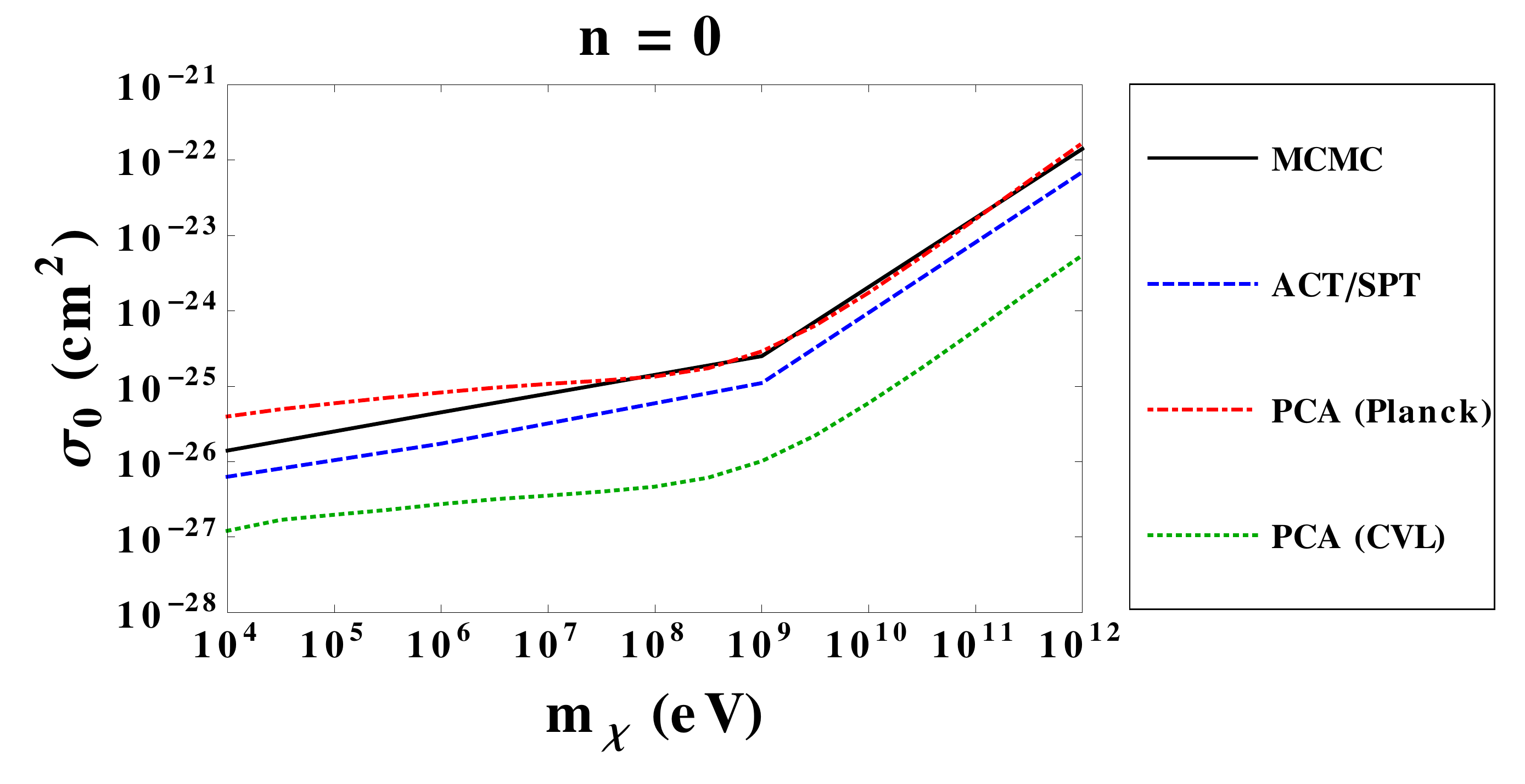}
     \includegraphics[width=9cm]{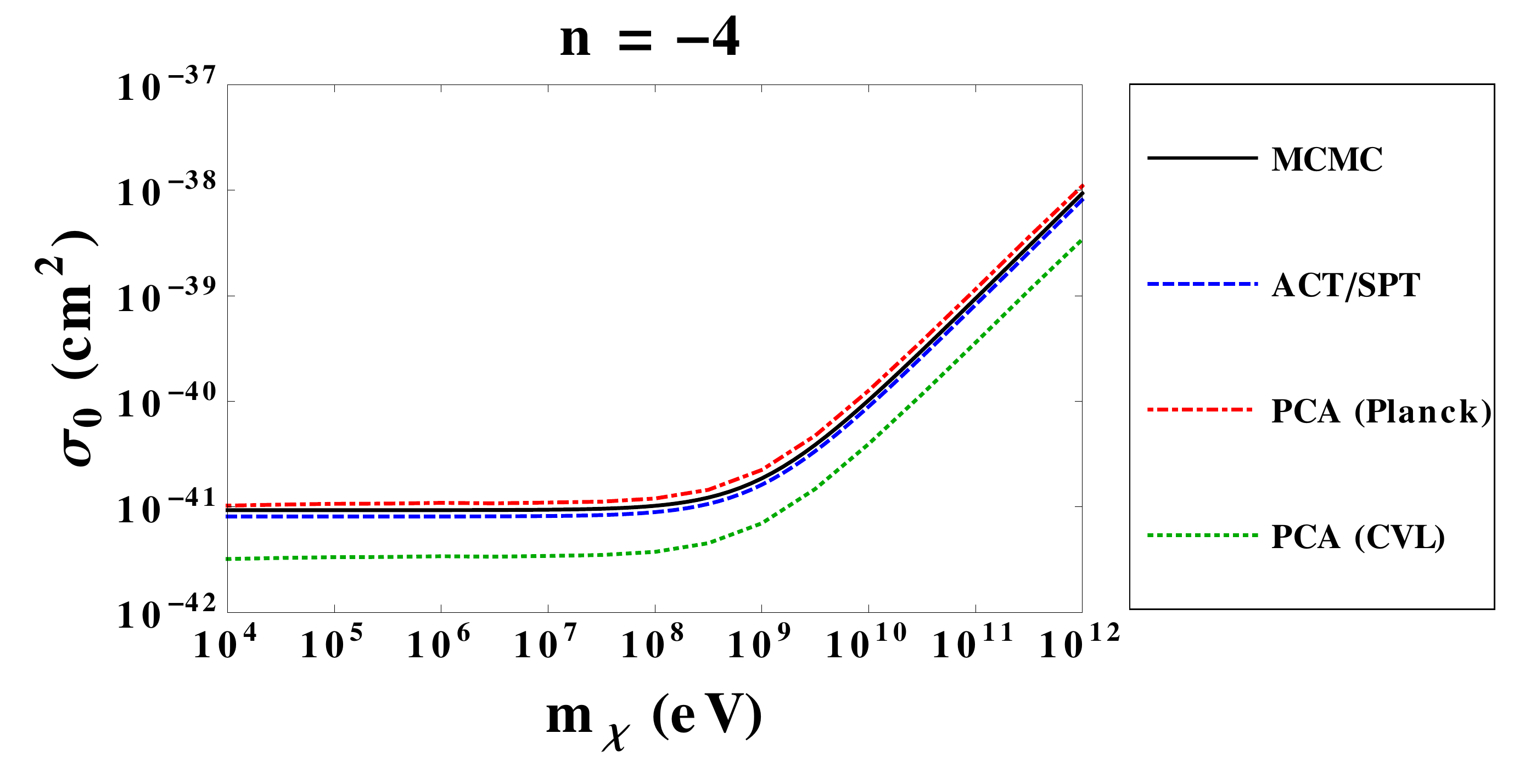}
  \caption{
   Comparison between the MCMC-based constraints (\emph{solid black line}; calculated with the \emph{Planck} likelihoods, and \emph{dashed blue line}; calculated with ACT/SPT) on the scattering cross section as a function of DM mass, for two different choices of the velocity dependence of scattering, with the forecast sensitivity from a Fisher analysis for \emph{Planck} sensitivity (\emph{dot-dashed red line}). We also show the predicted sensitivity of a future idealized experiment that is cosmic-variance limited (CVL) up to $\ell_\text{max} = 5000$ (\emph{dotted green line}).}
\label{pcaforecast}
\end{figure}

Using the Fisher-matrix formalism, we can estimate the predicted sensitivity of \emph{Planck} for arbitrary DM masses, once the velocity dependence of $\sigma$ is specified. In Fig.~\ref{pcaforecast} we show the results of this method and compare with the results of a full MCMC analysis; the agreement is good (within about 15\%) across the mass range we test, except for DM masses below 10 MeV for $n=0$ (where the estimate breaks down due to the nonlinearities we have discussed). We also show a Fisher forecast for an experiment with $f_\text{sky}$ comparable to \emph{Planck} that is cosmic variance limited (CVL) up to $\ell_\text{max} = 5000$; we see that we are currently within a factor of three of this limit for $n=-4$ scattering.

Because of the simple mass scaling, we can write the constraint on the scattering cross section for $n=-4$ to be:
\begin{align} \sigma_0 \lesssim \left( 1 + \frac{m_\chi}{m_H} \right) \begin{cases} 9.1 \times 10^{-42} \text{cm}^2 & \text{, \emph{Planck}} \\ 8.1 \times 10^{-42} \text{cm}^2 & \text{, \emph{Planck} + ACT/SPT} \\ 3.2 \times 10^{-42} \text{cm}^2 & \text{, CVL} \end{cases} \end{align}

\section{Modifying the thermal history at late times}
\label{sec:heating}

\subsection{Gas cooling in the cosmic dark ages and 21cm observations}

During reionization ($z\sim 6-10$) and at the end of the cosmic dark ages ($z\sim 10-200$), an important observable is the redshifted hydrogen hyperfine transition at 21cm wavelength (see Ref.~\cite{Furlanetto:2006jb} for a review). Measurements of 21cm radiation from the cosmic dark ages could potentially provide input to a number of important questions in cosmology, and can also be used to set constraints on DM-baryon scattering.

The size of the 21cm signal is controlled by the hydrogen spin temperature and the CMB temperature; if the former is smaller than the latter, the signal will be in absorption, whereas if the spin temperature exceeds the CMB temperature, an emission signal is expected. The spin temperature is expected to fall between the CMB temperature and the gas temperature; thus a measurement of an absorption trough sets an upper limit on the gas temperature, assuming the CMB temperature is known. (Conversely, measuring an emission peak would set a lower limit on the gas temperature, again assuming the CMB temperature was known.)

DM scattering with baryons could cool the hydrogen gas after the baryons decouple from the CMB radiation bath, thus lowering the gas temperature, enhancing 21cm absorption, and modifying the 21cm power spectrum \cite{Tashiro:2014tsa}. Ref.~\cite{Munoz:2015bca} showed that DM-baryon scattering can also heat both fluids under the right circumstances, from friction due to their relative velocity; this effect is more important for heavier DM, above 1 GeV in mass. 

Since any 21cm signal is expected to be sourced after the decoupling of the baryons and photons at $z\sim 150$ (at higher redshifts, the CMB and gas temperatures are identical, and no 21cm emission or absorption is expected), the relative velocity of DM and baryons is smaller than the velocities relevant for the CMB anisotropy constraints discussed above (which arise from the epoch prior to recombination). Thus we expect 21cm experiments to become increasingly sensitive, compared to the CMB anisotropy limits, for scattering that is enhanced at low velocity, and in particular for the $n=-4$ case.

\begin{figure}[h]
    \includegraphics[width=9cm]{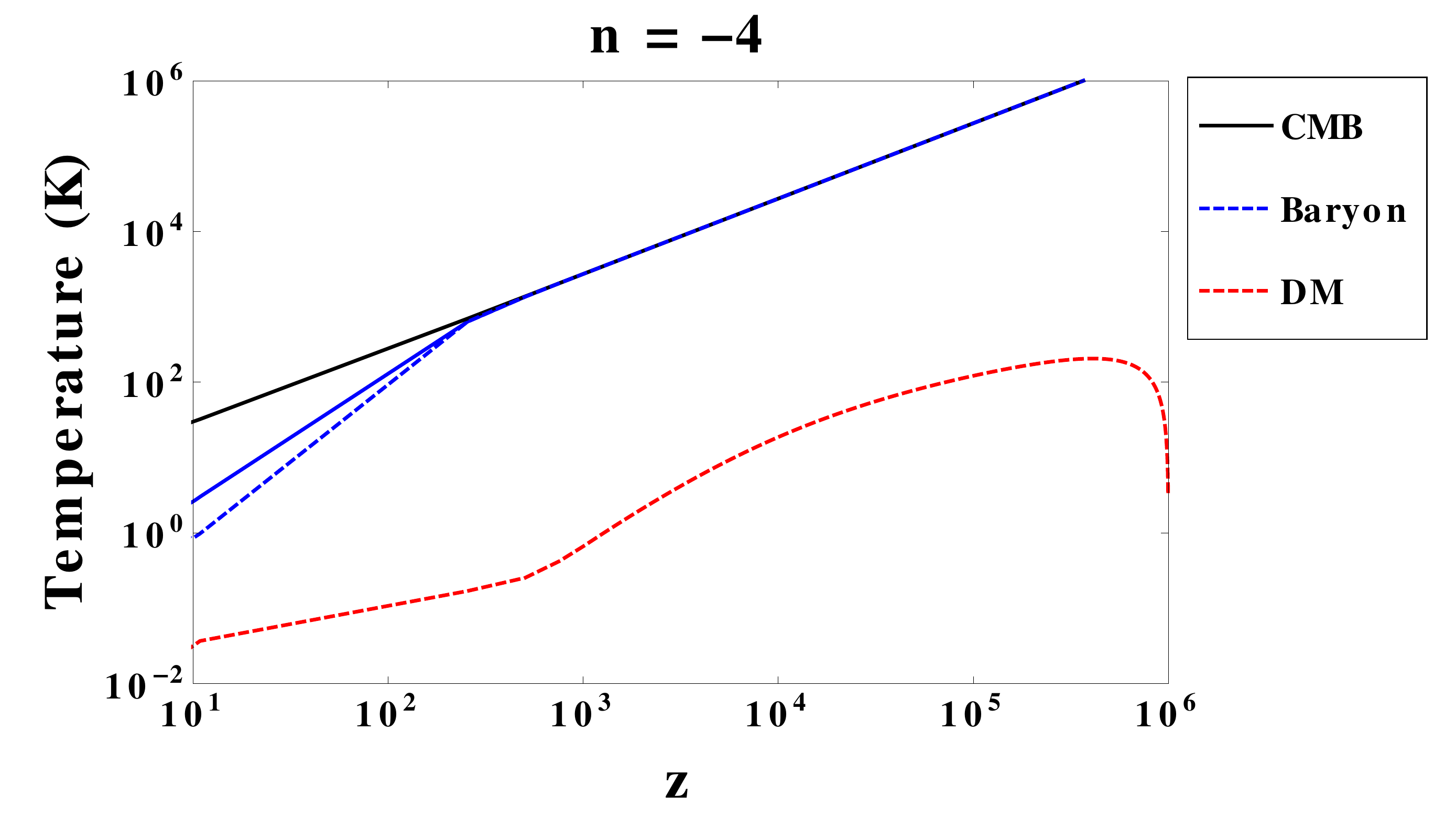}
  \caption{
   Example temperature evolution for baryons (\emph{blue}) and DM (\emph{red}) with (\emph{dashed}) and without (\emph{solid}) $n = -4$ DM-baryon scattering. The DM mass is taken to be 0.1 GeV, and the cross section is chosen to saturate the CMB-anisotropy limit derived in Sec.~\ref{sec:cmbconstraints}.}
\label{temp1}
\end{figure}

Assuming no initial DM-baryon relative bulk velocity at $z=10^6$, and using the mean-field approach for inclusion of the DM-baryon relative velocity (as for the anisotropy limits), an example of the evolution of the DM and baryon temperature is given in Fig.~\ref{temp1}.

\begin{figure}[h]
    \includegraphics[width=9cm]{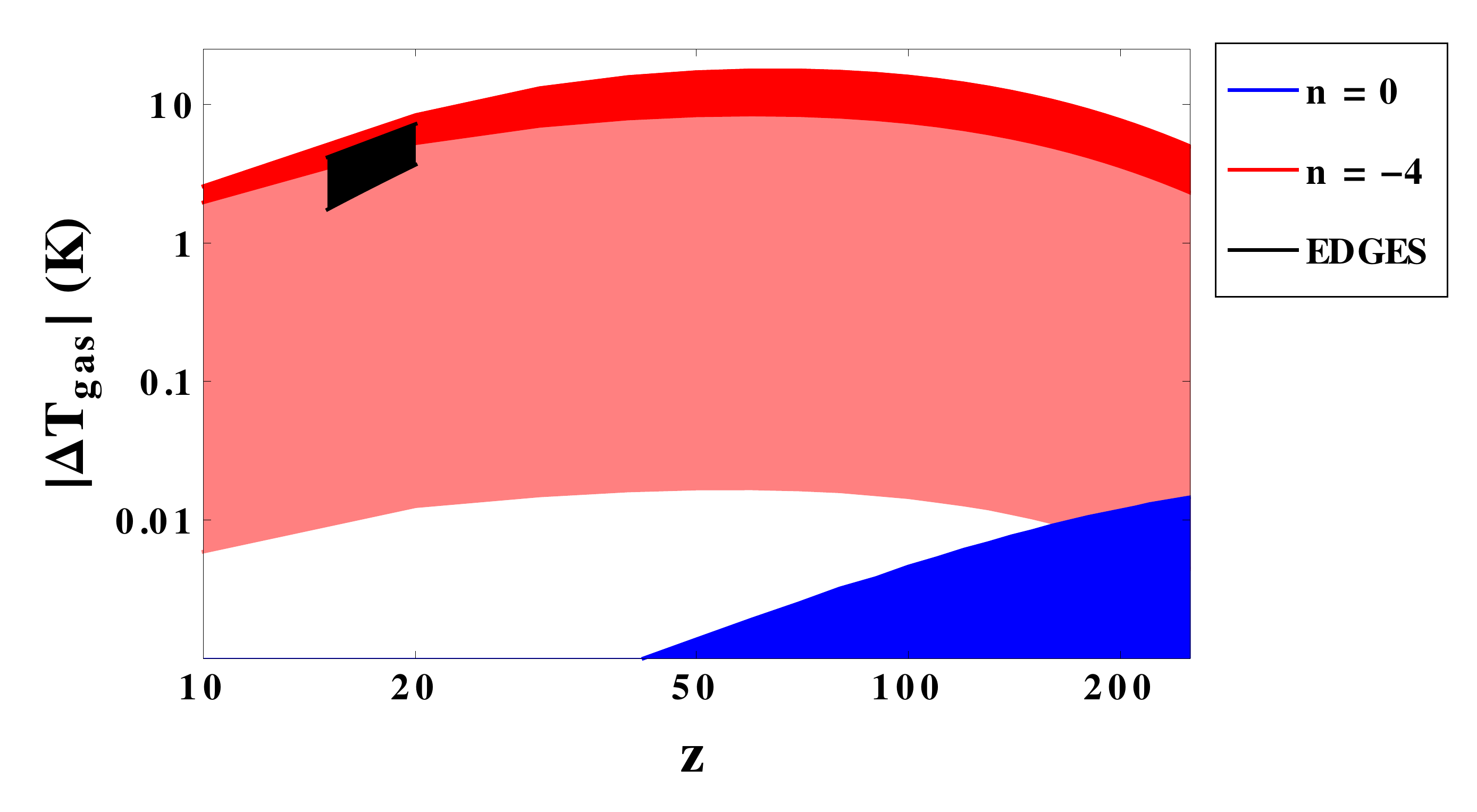}
  \caption{
  Maximum allowed change in baryon temperature due to DM-baryon scattering with $n = 0$ (\emph{blue}) and $n = -4$ (\emph{red}), scanning over keV-TeV DM masses. For each mass, the cross section is taken to saturate the CMB anisotropy limit derived in Sec.~\ref{sec:cmbconstraints} (using both \emph{Planck} and ACT/SPT data). The dark red band shows the range of maximum temperature modifications for sub-GeV DM masses, whereas the light red band covers the mass range up to 1 TeV. The black region shows the minimum gas temperature change, relative to the $\Lambda$CDM baseline, preferred by the recent measurement of EDGES  (95\% confidence region) \cite{bowman2018absorption}.}
\label{deltat}
\end{figure}

Focusing on low redshift, the maximum change in the baryon temperature with redshift, for different velocity scalings for the DM-baryon cross section, is shown in Fig.~\ref{deltat}. For $n =0$, the  change in the baryon temperature is very tiny (below $10^{-3}$ K at $z \lesssim 20$) assuming the maximum cross section consistent with the CMB-anisotropy limit (from \emph{Planck} + ACT/SPT) discussed above; only $n = -4$ (or stronger) scaling gives appreciable changes to the gas temperature at low redshift. The recent measurement of the Experiment to Detect the Reionization Step (EDGES) collaboration is also shown; this measurement indicates (assuming the only radiation background is the CMB) that baryons have a temperature of $T_b \leq 5.1 \,\text{K}$ at z = 17.2, which is lower than the standard cosmological model $T_b \sim 7 \,\text{K}$. The 21-cm signal $T_{21}$ in unit of mK is related to spin temperature $T_s$ by
\begin{equation}
T_{21} = 26.8 \, x_{H \Rmnum{1}}\, \dfrac{\rho_g}{\overline{\rho_g}} \left(\dfrac{1+z}{10}\right)\left(\dfrac{T_s-T_{CMB}}{T_s}\right) \,\,\,\text{mK},
\end{equation} 
where $x_{H \Rmnum{1}}$ is the mean mass fraction of neutral hydrogen, and $\rho_g$ and $\overline{\rho_g}$ respectively denote the gas density and its mean value. The error bar assumes no reionization, and saturated coupling such that $x_{H \Rmnum{1}} = 1$ and $T_s = T_\text{gas}$. We see that if 100\% of the DM is scattering, with a DM-baryon scattering cross section proportional to $v^{-4}$, then this process is just sufficient to match the best-fit EDGES result while remaining consistent with the CMB-anisotropy bounds, provided the DM is not much heavier than 1 GeV.

Given the proximity of our constraints to the cross sections needed to cool the baryons appreciably, one might worry about the effects of the imperfect modeling of the DM-baryon relative velocity. Going beyond the mean field approach in temperature evolution, Ref.~\cite{Munoz:2015bca} points out that the drag force between the DM and baryons could also heat up the baryons, and accounting for the evolution of the relative velocity between DM and baryons is important. The initial relative velocity $\bold{V}_{\chi b,0}$ at kinematic decoupling at $z \approx 1010$ follows a Gaussian distribution, 
\begin{equation}
P(\bold{V}_{\chi b,0}) = \dfrac{e^{-3 \bold{V}_{\chi b,0}^2/\left(2V_{rms}^2\right)}}{\left(\dfrac{2\pi}{3}V_{rms}^2\right)^{3/2}},
\end{equation}
where $V_{rms} \approx 29 \text{km/s}$ \cite{Ali-Haimoud:2013hpa}. The evolution equations of temperature and relative velocity are given in \cite{Munoz:2015bca}. 

\begin{figure}[h]
    \includegraphics[width=9cm]{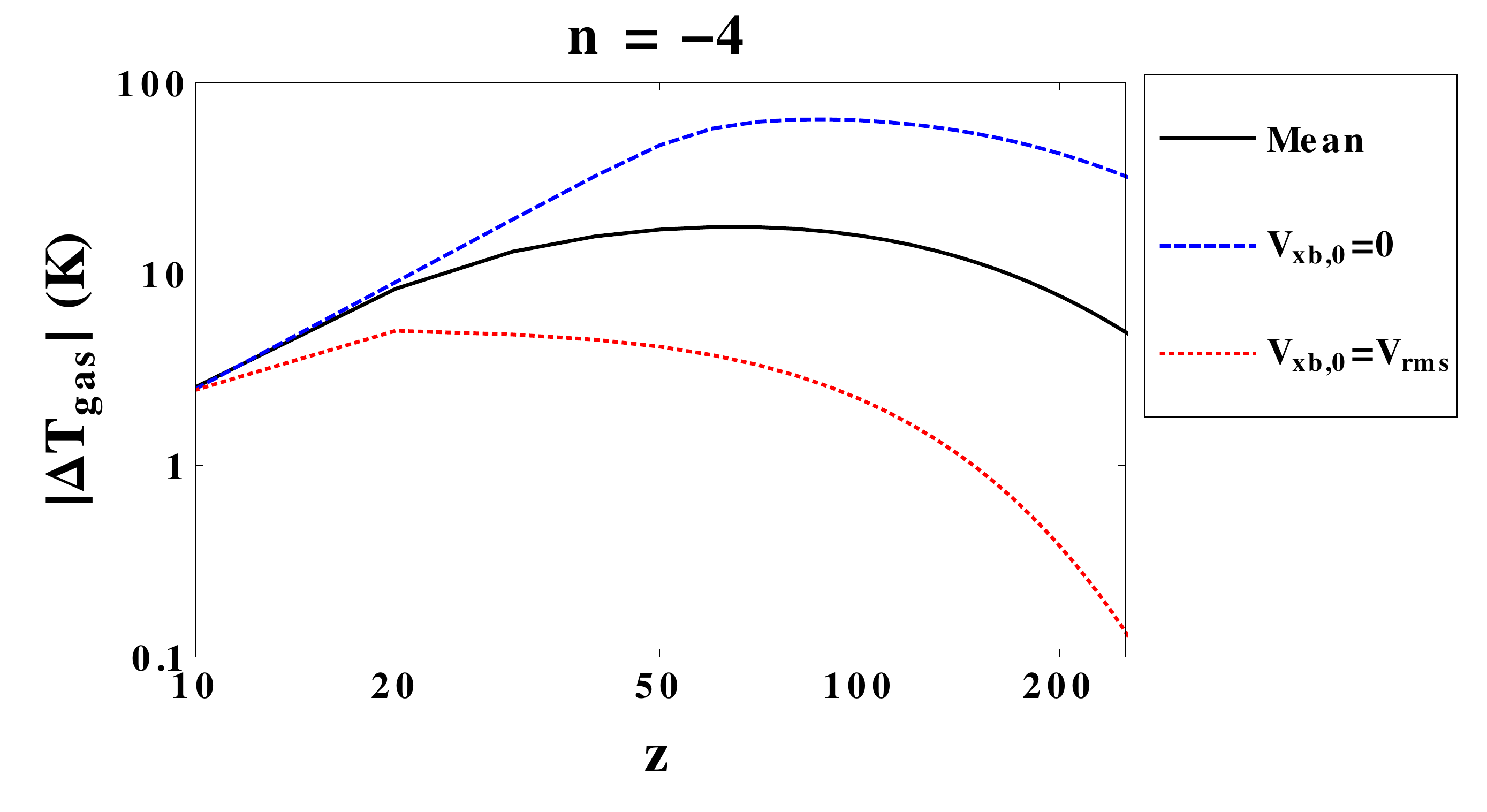}
  \caption{
  Change of baryon temperature with redshift for $n = -4$, for three different calculations of the DM-baryon relative velocity evolution: the mean-field approach (\emph{black solid}), and evolving $V_{\chi,b}$ with two initial conditions for $V_{\chi b,0}$ (\emph{blue dashed} and \emph{red dotted}). The DM mass is set to be 0.1 GeV, and the cross section saturates the CMB-anisotropy limit. }
\label{deltatvrms}
\end{figure}

Focusing on $n = -4$ scattering, we show in Fig.~\ref{deltatvrms} the resulting gas temperature evolution in the previous mean-field approach and with two initial conditions for $V_{\chi b,0}$, corresponding to $V_{\chi b,0} = 0, V_{rms}$. We see that the mean-field estimate interpolates between the two other cases, and in all three cases, few-K temperature changes can be achieved at $z\sim 17$ with the allowed cross sections.

\subsection{Constraints on late-time heating from the Lyman-alpha forest}

One might ask whether there is an additional lever arm on the DM-baryon scattering cross section from lower-redshift measurements of the gas temperature. In this section we review constraints on heating of the IGM after reionization, first calculated by Ref.~\cite{Munoz:2017qpy}.

After reionization, for redshifts $z \lesssim 7$, the processes that affect the temperature of the intergalactic medium (IGM) are described by \cite{Hui:1997dp,Sanderbeck:2015bba}:
\begin{equation}
\dot{T_b} = Q_\text{adia} + Q_\text{CMB} + Q_\text{ph} + Q_\text{cooling}.
\end{equation}
Here $Q_\text{adia}$ describes the temperature change due to expansion of universe, $Q_\text{CMB}$ describes the cooling/heating rate due to scattering on the CMB, $Q_\text{ph}$ is the photoheating rate of ionized hydrogen, and $Q_\text{cooling}$ includes recombination cooling, free-free cooling, and collisional cooling, etc. During $4 <  z < 7$, the gas temperature is roughly determined by the equilibrium between $Q_\text{ph}$,  $Q_\text{adia}$  and $Q_\text{CMB}$. At late times, when $z<4$, He \Rmnum{2} is ionized by X- ray emission from quasars, which can then raise the gas temperature; the presence of this He \Rmnum{2} process adds more uncertainty to the modeling of photoheating. Even for $z > 4$, photoheating of the intergalactic medium depends on the distribution of ionizing sources. 

\begin{figure}[h]
     \includegraphics[width=9cm]{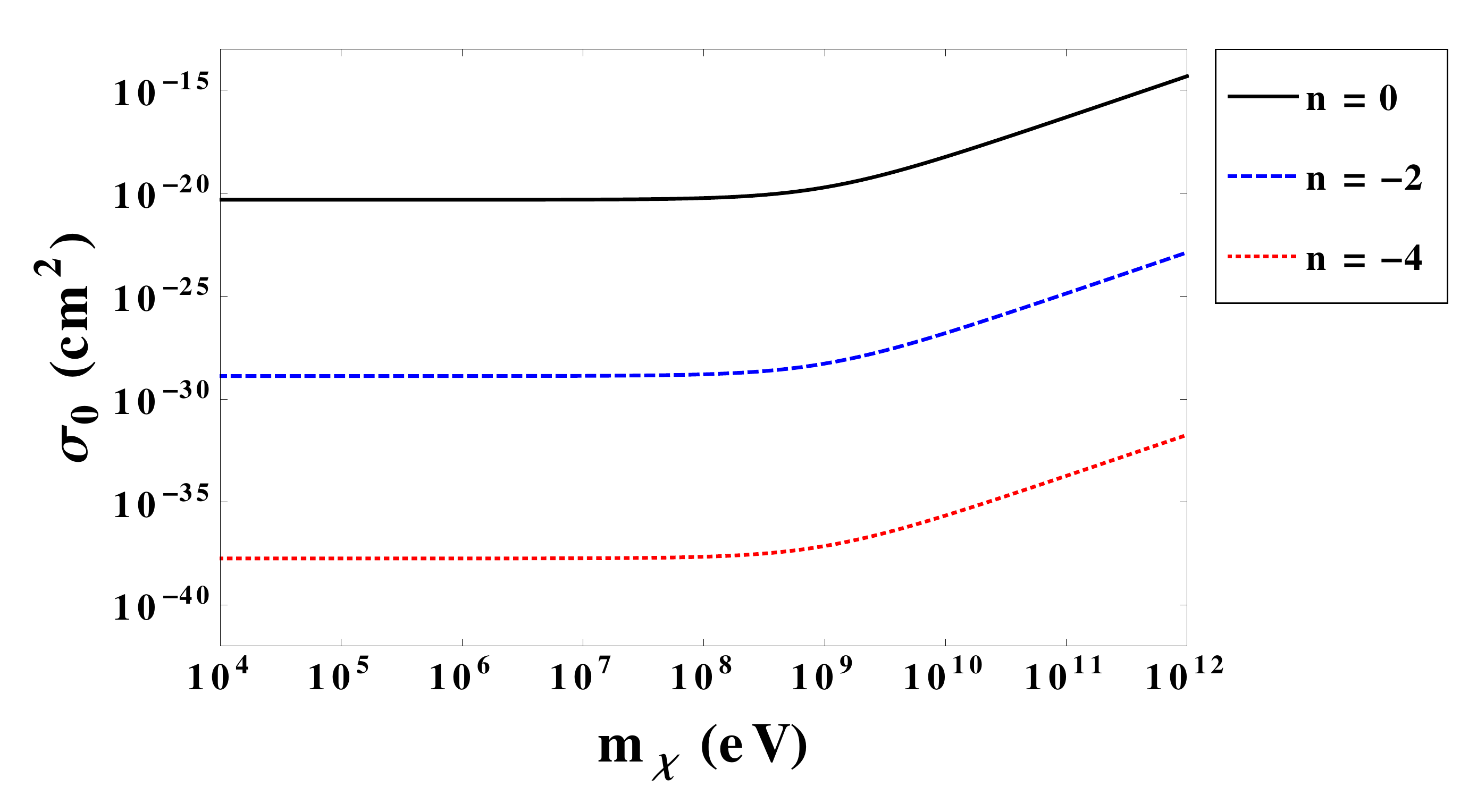}
  \caption{
   IGM limit for different velocity-dependent cross section assuming current sensitivity of Lyman-$\alpha$ forest data.}
\label{IGM}
\end{figure}

As discussed above, DM-baryon scattering can potentially cool the IGM, and so measurements of the IGM temperature at late times (in addition to the 21cm constraints discussed earlier) could provide a bound on the cross section \cite{Munoz:2017qpy}. Ref.~\cite{Munoz:2017qpy} found by numerical calculation that there is a roughly constant (with respect to redshift) shift in the gas temperature when $4 < z< 12$ in the presence of DM-baryon scattering, so the constraints can be estimated as: 
\begin{equation}
\Delta T_b = \dfrac{2}{3} \int dt \Gamma_{b,\chi} \left(T_\chi-T_b\right).
\end{equation}
Assuming $\Delta T_b/T_b < 0.1$, which is around the current sensitivity of Lyman-$\alpha$ forest data \cite{Viel:2013apy,McDonald:2004xn}, our calculated limit is shown in Fig.~\ref{IGM}. For the $n=0$ and $n=-4$ cases, our results agree well with Ref.~\cite{Munoz:2017qpy}.

 To further disentangle assumptions on the degree of photoheating vs the DM signal, we would need a deeper understanding of photoheating -- for example, the redshift dependence of photoheating. At present, these late-time heating constraints do not place a constraint on the scattering interpretation of the EDGES result.

\section{Spectral distortion}
\label{sec:distortion}

In addition to modifying the anisotropies of the CMB, DM-baryon scattering can also affect the overall blackbody spectrum. For $z < 2 \times 10^6$, photon-number-changing processes become inefficient, and injection of additional energy into (or from) the CMB will generically give rise to a distortion of the thermal blackbody spectrum \cite{Hu:1992dc}. For $z \gtrsim 5 \times 10^4$, Compton scattering efficiently redistributes photons in frequency, and the spectral distortion has the form of a chemical potential (described as a $\mu$-type distortion). At lower redshifts, the spectral shape of the distortion is modified; if the bath of electrons with which the CMB interacts is heated or cooled, this gives rise to a Compton-$y$ type distortion. Searches for $\mu$- and $y$-type distortions have been used to constrain energy injections from DM annihilations and decays \cite{Chluba:2013wsa,Hu:1993gc}, and the effects of DM-baryon scattering in Ref.~\cite{Ali-Haimoud:2015pwa}.

At low redshift, processes like the thermal Sunyaev-Zeldovich (SZ) effect and Compton scattering by free electrons in clusters are expected to produce a $y$-distortion comparable in size to the sensitivity of future experiments such as the proposed Primordial Inflation Explorer (PIXIE) \cite{Hill:2015tqa}. Given these non-trivial low-redshift backgrounds, we will focus on $\mu$-type distortions from early redshifts.

The fractional spectral distortion $\Delta$ can be estimated by the rate at which CMB photons change in energy due to Compton scattering, $\Delta =\Delta \rho_\gamma/\rho_\gamma$, with the evolution equation for this quantity being: 
\begin{equation}
\rho_\gamma \dfrac{d\Delta}{dt} = \dfrac{3}{2} n_b \dfrac{2 \mu_b}{m_e} R_\gamma \left( T_b-T_\gamma \right).
\end{equation} 
The baryon temperature is related to the DM-baryon scattering rate by Eq.~\ref{tempevol}, and assuming $T_b \approx T_\gamma$ at early times relevant for spectral distortions, we obtain:
\begin{equation}
\rho_\gamma \dfrac{d\Delta}{dt} = -\dfrac{3}{2} \left( N_b + \dfrac{2 \rho_\chi}{m_\chi+m_b} \dfrac{R_\chi \left(T_b - T_\chi \right) }{H T_b} \right) H T_\gamma.\label{eq:deltaev}
\end{equation} 
The authors of Ref.~\cite{Ali-Haimoud:2015pwa} forecast constraints from a PIXIE-like experiment on DM with mass in the keV-GeV range, using a simple analytic form for the spectral distortion from scattering. 
\begin{equation}
\Delta \approx -0.56 \left( \dfrac{n_b}{n_\gamma} \text{log}\left(\dfrac{10^{-4}}{5\times10^{-7}}\right)+\dfrac{n_\chi}{n_\gamma} \text{log}\left(\dfrac{a_{DB}}{5\times10^{-7}}\right) \right),
\end{equation}
where $n_\gamma$ is the number density of CMB photons, and $a_{DB}$ is a scale factor characterizing the time at which DM-baryon scattering decouples, with a cutoff for $\mu$-type distortion:
\begin{eqnarray}
a_{DB}=&\text{max} \Big\{10^{-4},&\text{min}\Big\{5\times10^{-7}, \nonumber \\
&&\dfrac{2}{3} \dfrac{2 m_\chi}{m_\chi+m_b} \dfrac{R_\chi \left(T_b - T_\chi \right) }{H T_b} \Big\}\Big\}.
\end{eqnarray}
However,  this analytic form is not a good approximation for $n<-2$, and consequently Ref.~\cite{Ali-Haimoud:2015pwa} did not present results for this case. For $n = -4$, for example, we instead need to solve for the spectral distortion numerically.\footnote{Note that we still treat the DM velocity distribution as being approximately Maxwellian for purposes of this estimate; a more detailed calculation would involve solving for the full evolution of the distribution.}

\begin{figure}[h]
    \includegraphics[width=9cm]{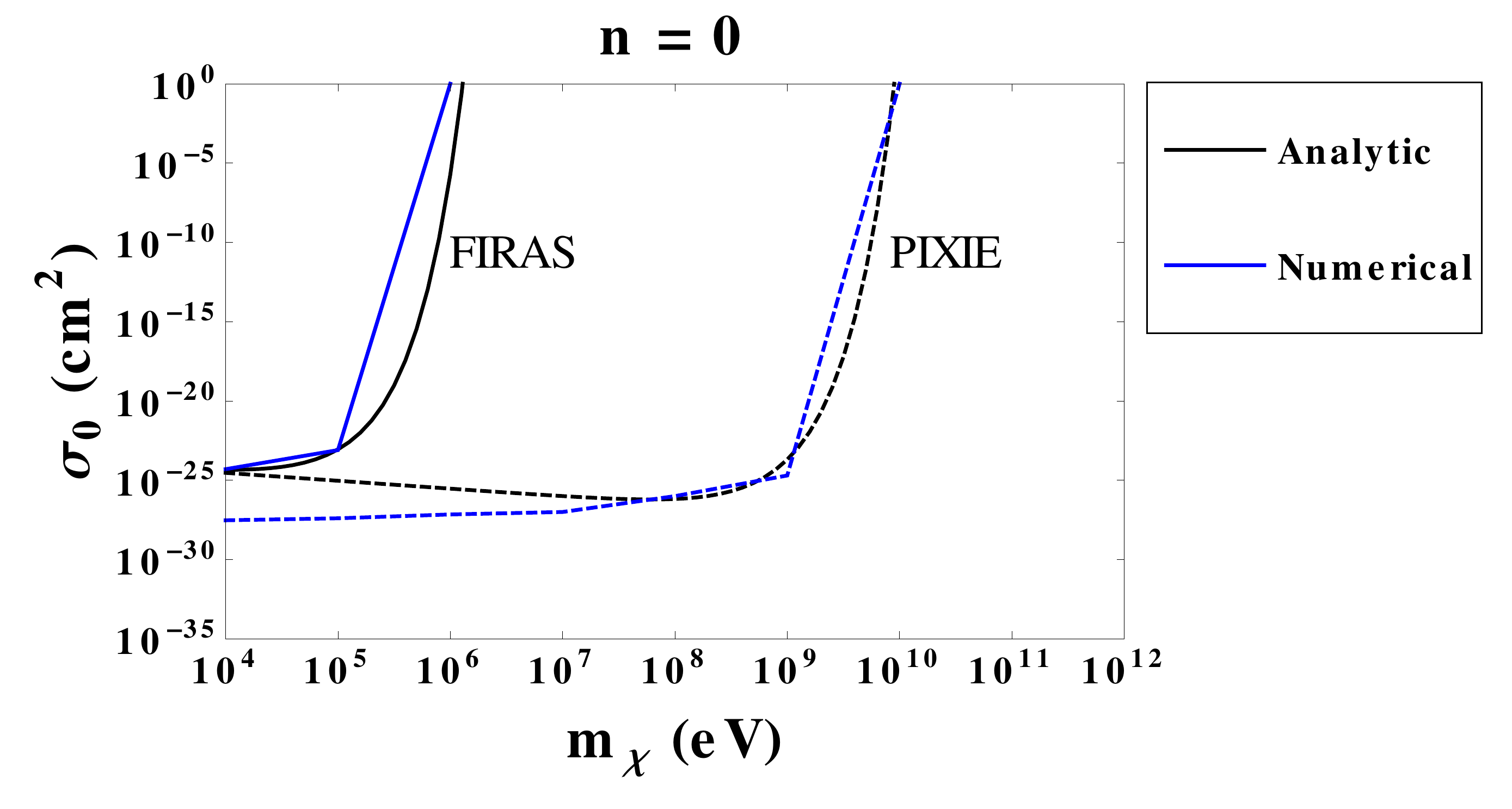}
   \includegraphics[width=9cm]{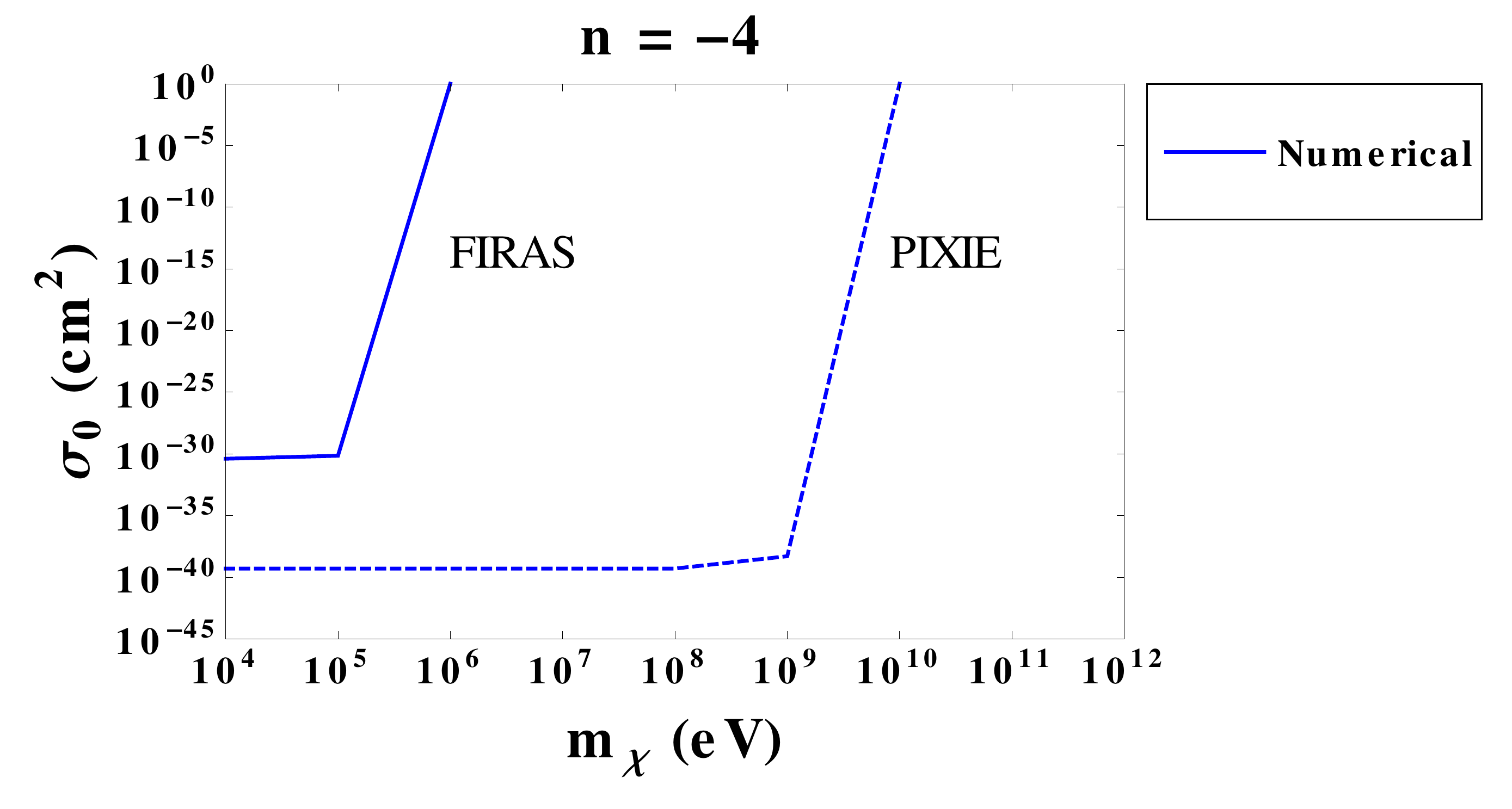}   
  \caption{
   Estimated cross section limit from spectral distortion constraints for $n = 0$ and $n = -4$ scattering, assuming FIRAS (\emph{solid line}) or PIXIE (\emph{dashed line}) sensitivity, using the analytic solution in \cite{Ali-Haimoud:2015pwa} (\emph{black}), and the numerical results of this work (\emph{blue}).}
\label{SD}
\end{figure}

We evolve Eq.~\ref{eq:deltaev} numerically from $z=10^7$ to $z=10^4$, using our previous results for the evolution of the DM and baryon temperatures, and then integrate over the range $z=10^4-2\times10^6$ to obtain $\Delta$. The Far-Infrared Absolute Spectrophotometer (FIRAS) has excluded $\Delta \gtrsim 5 \times 10^{-5}$ \cite{Fixsen:1996nj}, whereas a future PIXIE-like experiment \cite{Kogut:2011xw} could have sensitivity to $\Delta \sim 10^{-8}$ . The constraint from FIRAS and sensitivity estimate for a next-generation PIXIE-like experiment are shown in Fig.~\ref{SD}. We see that with FIRAS data the constraint cuts off quickly for DM mass scales above 100 keV -- 1 MeV, while PIXIE or a similar experiment could set stringent constraints up to 1 GeV DM masses.

This motivates us to consider the constraints on models where DM with a small electric millicharge constitutes some subdominant fraction of the total DM abundance, as suggested to explain the EDGES 21cm observation by Refs.~\cite{Munoz:2018pzp, Berlin:2018sjs, Barkana:2018qrx}. If the interacting DM fraction is below $\sim 1\%$, the effect on CMB anisotropies is small \cite{Dolgov:2013una}, as the behavior of the DM perturbations is governed primarily by the dominant non-interacting component. However, at high redshift, the interacting component of DM can scatter with charged particles (mostly protons, electrons and fully ionized helium) and still yield a non-negligible spectral distortion. The formalism for scattering with different target is similar to that laid out in Sec.~\ref{sec:formalism}, but replacing the reduced mass $\mu_{\chi, H}$ with the reduced mass $\mu_{\chi,t}$, where ``t'' denotes the scattering target and can represent either protons or electrons; scattering with helium is included as a modification to the proton-scattering term, as in Eq.~\ref{hecor}. The scattering cross section for a DM component with charge $\epsilon$ is \cite{McDermott:2010pa}:
\begin{equation}
\sigma = \dfrac{2 \pi \alpha^2 \epsilon^2 \xi}{\mu_{\chi,t}^2 v^4}
\end{equation}
where $\alpha$ is the fine structure constant and $\xi$ is the Debye logarithm, which can be approximated as:
\begin{equation}
\xi \approx 68 -2\, \text{log}\left(\dfrac{\epsilon}{10^{-6}}\right).
\end{equation}

  \begin{figure}[b]
    \includegraphics[width=9cm]{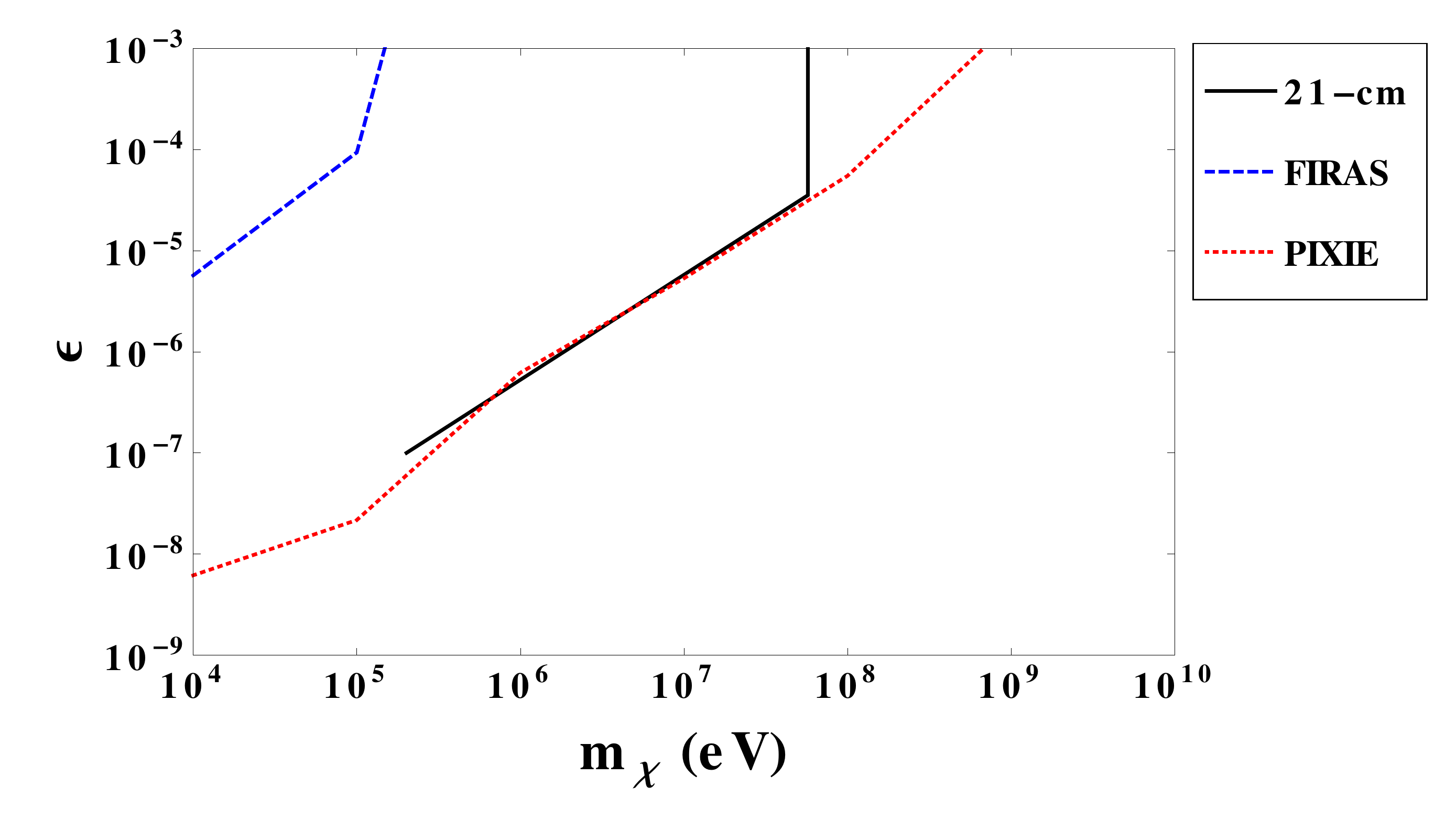}
  \caption{
   Preferred parameters for millicharged DM to explain the EDGES 21cm absorption detection \cite{Munoz:2018pzp}  (\emph{black line}) with $1\%$ of the DM being millicharged, and the estimated sensitivity of experiments to measure spectral distortion, calculated in this work. The \emph{dashed blue line} reflects current constraints (FIRAS) whereas the \emph{dotted red line} corresponds to a future PIXIE-like experiment with sensitivity to $10^{-8}$ distortions.}
\label{millicharge}
\end{figure}

More explicitly, when scattering on multiple species is present, we can generalize Eq.~\ref{tempevol} to write (for the $n=-4$ case):
\begin{align}
\dot{T_\chi} &= - 2 \dfrac{\dot{a}}{a} T_\chi + 2 m_\chi a c_{-4} \left( T_b - T_\chi \right) \sum_t \frac{\rho_t \sigma_{0,t}}{(m_\chi +m_t)^2} \frac{1}{u_t^3}, \nonumber \\
\dot{T_b} &= - 2 \dfrac{\dot{a}}{a} T_b + 2\dfrac{\mu_b}{m_e} R_\gamma \left( T_\gamma - T_b \right)  \nonumber \\
& + \dfrac{2 a c_{-4} \rho_\chi}{n_H (1 + f_\text{He} + x_e)} \left( T_\chi - T_b \right) \sum_t \frac{\rho_t \sigma_{0,t}}{(m_\chi + m_t)^2} \frac{1}{u_t^3},
\end{align}
where the sum over $t$ describes different interacting species with mass $m_t$ and mass density $\rho_t$, $\sigma_{0,t}$ is the cross section $\sigma_0$ for the interactions of species $t$ with the baryons, and $u_t = (T_b/m_t + T_\chi/m_\chi + V_\text{rms}^2/3)^{1/2}$ (within the mean-field approach). Here $\rho_\chi$ should be taken to be the density of the DM that scatters with the baryons, and we assume there are no interactions between this component and the remainder of the DM.

We add together the contributions from electron-DM, proton-DM scattering and helium-DM scattering, assuming full ionization (as is valid for $z > 10^4$). The resulting constraint on the millicharge is shown in Fig.~\ref{millicharge}. As shown, a future experiment with sensitivity to spectral distortions at the $\mathcal{O}(10^{-8})$ level is capable of reaching this parameter space.

\section{Conclusion}
\label{sec:conclusion}
In this work, we have studied constraints on DM-baryon scattering with a velocity-dependent cross section, $\sigma = \sigma_0 v^n$ for $n \leq 0$, from CMB anisotropies, CMB spectral distortion, and the IGM temperature, and discussed implications for the global 21cm signal. We have shown that inclusion of ACT/SPT high-$\ell$ data improves the limit from the CMB anisotropies by about $20\%$ compared to \emph{Planck}-only results for scattering with $\sigma \propto v^{-4}$, but the recent measurement of 21-cm absorption by EDGES remains just consistent with the hypothesis that $100\%$ of the DM scatters on baryons with a cross section proportional to $v^{-4}$ (before accounting for model-specific considerations; there are independent constraints on the possibility that e.g. 100\% of the DM is millicharged, or interacts through a new low-mass dark photon). We have mapped out the redshifts that dominate the CMB anisotropy constraints, and demonstrated quantitatively that for $n=-4$ scattering (a) the problem is approximately linear, and (b) the mass dependence of the forecast limit can be captured by a single simple parameter, in agreement with observations in the literature. We have provided a principal component basis for estimating constraints on modified scattering histories. We have discussed the equivalent results for $n=0$ scattering, where a similar formalism can be applied for DM with mass 100 MeV and greater, but the problem becomes increasingly non-linear (due to interplay between the DM temperature and scattering cross section) for lighter DM masses.

We have demonstrated that future measurements of CMB spectral distortion have the potential to strongly constrain scattering with $\sigma \propto v^{-4}$, for DM masses below 1 GeV, even if the interacting component is only a small fraction of the total DM abundance.  We estimate that a future PIXIE-like experiment has the potential to test the hypothesis that the EDGES absorption signal results from scattering by a subdominant component of millicharged DM.

{\bf Acknowledgments:} We thank Rebecca Leane, Hongwan Liu, Nicholas Rodd, and Yotam Soreq for helpful discussions and feedback. We are particularly grateful to Cora Dvorkin and Weishuang Linda Xu, for providing clear, prompt and detailed information on their work, allowing us to cross-check our results; and to Kimberly Boddy and Vera Gluscevic, for calling our attention to an issue with the low-mass scaling of our limit in the $n=0$ case, in a preliminary draft of this work. This work was supported by the Office of High Energy Physics of the U.S. Department of Energy under grant Contract Numbers DE-SC00012567 and DE-SC0013999. Wu is partially supported by the Taiwan Top University Strategic Alliance (TUSA) Fellowship. 

\appendix

\section{Estimating CMB constraints on arbitrary redshift-dependent scattering histories}
\label{app:fourpcs}

As discussed in Sec.~\ref{sec:redshift}, when we perform a principal component analysis using a basis of histories where scattering occurs only for a short time, the first four principal components account for roughly $90\%$ of the variance, for both the $n=0$ and $n=-4$ cases. Thus for scattering histories broadly resembling either the $n=0$ or $n=-4$ cases, we can estimate the likely constraint on DM-baryon scattering from the CMB (provided, as discussed in Sec.~\ref{sec:linearity}, that the DM does not become hot enough that its thermal velocity dominates the DM-baryon relative velocity).

We provide in supplementary data files (\texttt{pca$\_$n=-4.dat} and \texttt{pca$\_$n=0.dat}) the list of coefficients $\alpha_{ij}$ such that the $i$th modulation function $G_i(z) = \sum_{j=1}^N \alpha_{ij} P_j(z)$, where $P_j(z)$ is the $j$th principal component and $N$ is the number of unit-normalized Gaussian basis functions. In the files, the first column gives the redshift $z_i$; the next four columns provide $\alpha_{i1}$, $\alpha_{i2}$, $\alpha_{i3}$, $\alpha_{i4}$ respectively.

Given a redshift-dependent cross section $\sigma(z)$, we can approximate:
\begin{equation} \sigma(z) \approx \sum_i \sigma(z_i) G_i(z) \approx \frac{\sigma(z_i)}{\sigma_0 v(z_i)^{-n}} \sigma_0 v(z)^{n} G_i(z), \end{equation}
since $G_i(z)$ approximates a narrow step function covering the range $z_i \pm (\Delta z_i)/2$. 

  \begin{table}[h]
  \def\arraystretch{1.3}
    \begin{tabular}{|l|l|l|}
\hline
& $n=0$ & $n=-4$ \\
\hline
$\sigma_0$ (cm$^2$) & $10^{-26}$ & $10^{-42}$ \\
\hline
$\lambda_1 \times 100$ & 9.8 & 7.8 \\
$\lambda_2 \times 100 $ & 7.4 & 6.4 \\
$\lambda_3 \times 100 $ & 2.1 & 3.2 \\
$\lambda_4 \times 100 $ & 1.6 & 1.8 \\
\hline
\end{tabular}
  \caption{Fisher-matrix eigenvalues $\lambda_i$ for the first four principal components, ranked in order of eigenvalue, calculated with a baseline scattering history given by $\sigma = \sigma_0$ (middle column) and $\sigma = \sigma_0 v^{-4}$ (right column), for 0.1 GeV DM. Values of the baseline cross section $\sigma_0$ are given in the second row. These results are for a \emph{Planck}-like experiment.
  }
\label{tab:eigenvalues}
\end{table}

Here $\sigma = \sigma_0 v^n$ is the ``baseline'' scattering history which multiplies the $G_i(z)$ modulation factors, in calculating the Fisher matrix and the principal components. It can be translated into a redshift-dependent scattering history by approximating $T_b \approx T_\text{CMB}(z) = (1+z) T_\text{CMB,0}$ and writing:
\begin{align} v^n & \rightarrow c_n \left(\frac{T_b}{m_H} + \frac{V_\text{rms}^2}{3} \right)^{(n+1)/2} \nonumber \\
& = c_n \left(\frac{T_\text{CMB,0} (1+z)}{m_H} + \frac{V_\text{rms}^2}{3} \right)^{(n+1)/2}. \end{align}
The normalizations $\sigma_0$ and the corresponding eigenvalues of the first four principal components are given in Table~\ref{tab:eigenvalues}.

Thus in terms of the $\alpha_{ij}$ coefficients, we can write:
\begin{align} \sigma(z) & \approx \sum_j \left[P_j(z) \sigma_0 c_n \left(\frac{T_\text{CMB,0} (1+z)}{m_H} + \frac{V_\text{rms}^2}{3} \right)^{(n+1)/2} \right] \nonumber \\
& \times \sum_i \alpha_{ij} \frac{\sigma(z_i)}{\sigma_0 c_n \left(\frac{T_\text{CMB,0} (1+z_i)}{m_H} + \frac{V_\text{rms}^2}{3} \right)^{(n+1)/2}}.\end{align}
The term in square brackets describes the physical scattering history corresponding to the $j$th principal component; by construction, these histories have approximately orthogonal effects on the CMB after marginalization over the cosmological parameters, and so the significances of the corresponding signals add in quadrature. The significance (in sigma) of the $i$th such orthogonal perturbation can be estimated as $\sqrt{\lambda_i}$, where $\lambda_i$ is the Fisher-matrix eigenvalue corresponding to the $i$th principal component. Thus the number of sigma at which the $\sigma(z)$ history could be detected (in this case, by a \emph{Planck}-like experiment) can be estimated as:
\begin{equation} \sqrt{\sum_j \lambda_j \left(\sum_i \alpha_{ij} \frac{\sigma(z_i)}{\sigma_0 c_n \left(\frac{T_\text{CMB,0} (1+z)}{m_H} + \frac{V_\text{rms}^2}{3} \right)^{(n+1)/2}} \right)^2}.\end{equation}

A constraint can then be placed on the normalization of the scattering history $\sigma(z)$ by requiring that e.g.
\begin{equation} \sqrt{\sum_j \lambda_j \left(\sum_i \alpha_{ij} \frac{\sigma(z_i)}{c_n \left(\frac{T_\text{CMB,0} (1+z)}{m_H} + \frac{V_\text{rms}^2}{3} \right)^{(n+1)/2}} \right)^2} < 2.\end{equation}
corresponding to the requirement that the level of the signal in \emph{Planck} data would be less than 2 sigma. 

The approximation of a truncated principal component analysis is to only include the first few terms in the sum over $j$, justified if $\lambda_j$ decreases rapidly with increasing $j$. This approximation can break down if the coefficients of $\lambda_j$ also vary strongly with $j$, and e.g. are suppressed for small $j$ / enhanced for large $j$. This is most likely to occur if $\sigma(z_i)$ is very different from the baseline scattering history, where the ratio $\sigma(z_i)/(\sigma_0 v(z_i)^n)$ can become very large; accordingly, we provide the $\lambda_j$ and $\alpha_{ij}$ coefficients for both the $n=0$ and $n=-4$ baselines, so that whichever is more similar to the desired scattering scenario can be used. In general, this Fisher-matrix-based approach may also break down due to non-Gaussianity of the likelihood, or a breakdown of the assumption of linearity (i.e. that the effect of a sum of scattering histories on the CMB is the same as the sum of their individual effects on the CMB). Note that while truncating the series at a smaller number of principal components always \emph{decreases} the calculated significance, this is not true for these other sources of error, and they can lead to a too-strong apparent limit.

As a simple example, we can test the case $\sigma = \sigma_{-2,0} v^{-2}$. Using the first four PCs associated with either the $n=0$ or $n=-4$ basis, our estimated 2-sigma constraint on $\sigma_{-2,0}$ becomes:
\begin{equation} \sigma_{-2,0} < \frac{2 \sigma_0 c_n/c_{-2}}{ \sqrt{\sum_{j=1}^4 \lambda_j \left(\sum_i \alpha_{ij}  \left(\frac{T_\text{CMB,0} (1+z_i)}{m_H} + \frac{V_\text{rms}^2}{3} \right)^{-1 - n/2}\right)^2}}\end{equation}

For $n=-4$, we obtain $\sigma_{-2,0} \lesssim 1.3 \times 10^{-33}$ cm$^2$. This agrees well with the results of a Fisher-matrix analysis, and with the results of \cite{Xu:2018efh}. For $n=0$, we obtain a somewhat weaker constraint, $\sigma_{-2,0} \lesssim 3.3 \times 10^{-33}$ cm$^2$ (likely due to the first few $n=0$ PCs not adequately capturing the increase in the scattering rate at smaller $z$).

The results given in this section and the supplemental material are for 0.1 GeV DM, but as discussed in Sec.~\ref{sec:mass}, for other masses the limiting cross section can simply be rescaled by a factor $(1 + m_\chi/m_H)$, to a good approximation, provided the DM temperature does not grow sufficiently large to influence the DM-baryon relative velocity.

\bibliography{decaypca}

\begin{thebibliography}{54}
\expandafter\ifx\csname natexlab\endcsname\relax\def\natexlab#1{#1}\fi
\expandafter\ifx\csname bibnamefont\endcsname\relax
  \def\bibnamefont#1{#1}\fi
\expandafter\ifx\csname bibfnamefont\endcsname\relax
  \def\bibfnamefont#1{#1}\fi
\expandafter\ifx\csname citenamefont\endcsname\relax
  \def\citenamefont#1{#1}\fi
\expandafter\ifx\csname url\endcsname\relax
  \def\url#1{\texttt{#1}}\fi
\expandafter\ifx\csname urlprefix\endcsname\relax\def\urlprefix{URL }\fi
\providecommand{\bibinfo}[2]{#2}
\providecommand{\eprint}[2][]{\url{#2}}

\bibitem[{\citenamefont{Dvorkin et~al.}(2014)\citenamefont{Dvorkin, Blum, and
  Kamionkowski}}]{Dvorkin:2013cea}
\bibinfo{author}{\bibfnamefont{C.}~\bibnamefont{Dvorkin}},
  \bibinfo{author}{\bibfnamefont{K.}~\bibnamefont{Blum}}, \bibnamefont{and}
  \bibinfo{author}{\bibfnamefont{M.}~\bibnamefont{Kamionkowski}},
  \bibinfo{journal}{Phys. Rev.} \textbf{\bibinfo{volume}{D89}},
  \bibinfo{pages}{023519} (\bibinfo{year}{2014}), \eprint{1311.2937}.

\bibitem[{\citenamefont{Gluscevic and Boddy}(2017)}]{Gluscevic:2017ywp}
\bibinfo{author}{\bibfnamefont{V.}~\bibnamefont{Gluscevic}} \bibnamefont{and}
  \bibinfo{author}{\bibfnamefont{K.~K.} \bibnamefont{Boddy}}
  (\bibinfo{year}{2017}), \eprint{1712.07133}.

\bibitem[{\citenamefont{Boddy and Gluscevic}(2018)}]{Boddy:2018kfv}
\bibinfo{author}{\bibfnamefont{K.~K.} \bibnamefont{Boddy}} \bibnamefont{and}
  \bibinfo{author}{\bibfnamefont{V.}~\bibnamefont{Gluscevic}}
  (\bibinfo{year}{2018}), \eprint{1801.08609}.

\bibitem[{\citenamefont{Xu et~al.}(2018)\citenamefont{Xu, Dvorkin, and
  Chael}}]{Xu:2018efh}
\bibinfo{author}{\bibfnamefont{W.~L.} \bibnamefont{Xu}},
  \bibinfo{author}{\bibfnamefont{C.}~\bibnamefont{Dvorkin}}, \bibnamefont{and}
  \bibinfo{author}{\bibfnamefont{A.}~\bibnamefont{Chael}}
  (\bibinfo{year}{2018}), \eprint{1802.06788}.

\bibitem[{\citenamefont{Ali-Haïmoud et~al.}(2015)\citenamefont{Ali-Haïmoud,
  Chluba, and Kamionkowski}}]{Ali-Haimoud:2015pwa}
\bibinfo{author}{\bibfnamefont{Y.}~\bibnamefont{Ali-Haïmoud}},
  \bibinfo{author}{\bibfnamefont{J.}~\bibnamefont{Chluba}}, \bibnamefont{and}
  \bibinfo{author}{\bibfnamefont{M.}~\bibnamefont{Kamionkowski}},
  \bibinfo{journal}{Phys. Rev. Lett.} \textbf{\bibinfo{volume}{115}},
  \bibinfo{pages}{071304} (\bibinfo{year}{2015}), \eprint{1506.04745}.

\bibitem[{\citenamefont{Bowman et~al.}(2018)\citenamefont{Bowman, Rogers,
  Monsalve, Mozdzen, and Mahesh}}]{bowman2018absorption}
\bibinfo{author}{\bibfnamefont{J.~D.} \bibnamefont{Bowman}},
  \bibinfo{author}{\bibfnamefont{A.~E.} \bibnamefont{Rogers}},
  \bibinfo{author}{\bibfnamefont{R.~A.} \bibnamefont{Monsalve}},
  \bibinfo{author}{\bibfnamefont{T.~J.} \bibnamefont{Mozdzen}},
  \bibnamefont{and} \bibinfo{author}{\bibfnamefont{N.}~\bibnamefont{Mahesh}},
  \bibinfo{journal}{Nature} \textbf{\bibinfo{volume}{555}}, \bibinfo{pages}{67}
  (\bibinfo{year}{2018}).

\bibitem[{\citenamefont{Barkana}(2018)}]{barkana2018possible}
\bibinfo{author}{\bibfnamefont{R.}~\bibnamefont{Barkana}},
  \bibinfo{journal}{Nature} \textbf{\bibinfo{volume}{555}}, \bibinfo{pages}{71}
  (\bibinfo{year}{2018}).

\bibitem[{\citenamefont{Mu–oz and Loeb}(2018)}]{Munoz:2018pzp}
\bibinfo{author}{\bibfnamefont{J.~B.} \bibnamefont{Mu–oz}} \bibnamefont{and}
  \bibinfo{author}{\bibfnamefont{A.}~\bibnamefont{Loeb}}
  (\bibinfo{year}{2018}), \eprint{1802.10094}.

\bibitem[{\citenamefont{Berlin et~al.}(2018)\citenamefont{Berlin, Hooper,
  Krnjaic, and McDermott}}]{Berlin:2018sjs}
\bibinfo{author}{\bibfnamefont{A.}~\bibnamefont{Berlin}},
  \bibinfo{author}{\bibfnamefont{D.}~\bibnamefont{Hooper}},
  \bibinfo{author}{\bibfnamefont{G.}~\bibnamefont{Krnjaic}}, \bibnamefont{and}
  \bibinfo{author}{\bibfnamefont{S.~D.} \bibnamefont{McDermott}}
  (\bibinfo{year}{2018}), \eprint{1803.02804}.

\bibitem[{\citenamefont{Barkana et~al.}(2018)\citenamefont{Barkana,
  Outmezguine, Redigolo, and Volansky}}]{Barkana:2018qrx}
\bibinfo{author}{\bibfnamefont{R.}~\bibnamefont{Barkana}},
  \bibinfo{author}{\bibfnamefont{N.~J.} \bibnamefont{Outmezguine}},
  \bibinfo{author}{\bibfnamefont{D.}~\bibnamefont{Redigolo}}, \bibnamefont{and}
  \bibinfo{author}{\bibfnamefont{T.}~\bibnamefont{Volansky}}
  (\bibinfo{year}{2018}), \eprint{1803.03091}.

\bibitem[{\citenamefont{Feng and Holder}(2018)}]{Feng:2018rje}
\bibinfo{author}{\bibfnamefont{C.}~\bibnamefont{Feng}} \bibnamefont{and}
  \bibinfo{author}{\bibfnamefont{G.}~\bibnamefont{Holder}}
  (\bibinfo{year}{2018}), \eprint{1802.07432}.

\bibitem[{\citenamefont{Fraser et~al.}(2018)}]{Fraser:2018acy}
\bibinfo{author}{\bibfnamefont{S.}~\bibnamefont{Fraser}} \bibnamefont{et~al.}
  (\bibinfo{year}{2018}), \eprint{1803.03245}.

\bibitem[{\citenamefont{Pospelov et~al.}(2018)\citenamefont{Pospelov, Pradler,
  Ruderman, and Urbano}}]{Pospelov:2018kdh}
\bibinfo{author}{\bibfnamefont{M.}~\bibnamefont{Pospelov}},
  \bibinfo{author}{\bibfnamefont{J.}~\bibnamefont{Pradler}},
  \bibinfo{author}{\bibfnamefont{J.~T.} \bibnamefont{Ruderman}},
  \bibnamefont{and} \bibinfo{author}{\bibfnamefont{A.}~\bibnamefont{Urbano}}
  (\bibinfo{year}{2018}), \eprint{1803.07048}.

\bibitem[{\citenamefont{Bernal et~al.}(2017)\citenamefont{Bernal, Raccanelli,
  Verde, and Silk}}]{Bernal:2017nec}
\bibinfo{author}{\bibfnamefont{J.~L.} \bibnamefont{Bernal}},
  \bibinfo{author}{\bibfnamefont{A.}~\bibnamefont{Raccanelli}},
  \bibinfo{author}{\bibfnamefont{L.}~\bibnamefont{Verde}}, \bibnamefont{and}
  \bibinfo{author}{\bibfnamefont{J.}~\bibnamefont{Silk}}
  (\bibinfo{year}{2017}), \eprint{1712.01311}.

\bibitem[{\citenamefont{Ewall-Wice et~al.}(2018)\citenamefont{Ewall-Wice,
  Chang, Lazio, Dore, Seiffert, and Monsalve}}]{Ewall-Wice:2018bzf}
\bibinfo{author}{\bibfnamefont{A.}~\bibnamefont{Ewall-Wice}},
  \bibinfo{author}{\bibfnamefont{T.~C.} \bibnamefont{Chang}},
  \bibinfo{author}{\bibfnamefont{J.}~\bibnamefont{Lazio}},
  \bibinfo{author}{\bibfnamefont{O.}~\bibnamefont{Dore}},
  \bibinfo{author}{\bibfnamefont{M.}~\bibnamefont{Seiffert}}, \bibnamefont{and}
  \bibinfo{author}{\bibfnamefont{R.~A.} \bibnamefont{Monsalve}}
  (\bibinfo{year}{2018}), \eprint{1803.01815}.

\bibitem[{\citenamefont{Liu and Slatyer}(2018)}]{hongwantoappear}
\bibinfo{author}{\bibfnamefont{H.}~\bibnamefont{Liu}} \bibnamefont{and}
  \bibinfo{author}{\bibfnamefont{T.~R.} \bibnamefont{Slatyer}},
  \bibinfo{journal}{to appear}  (\bibinfo{year}{2018}).

\bibitem[{\citenamefont{Sigurdson et~al.}(2004)\citenamefont{Sigurdson, Doran,
  Kurylov, Caldwell, and Kamionkowski}}]{Sigurdson:2004zp}
\bibinfo{author}{\bibfnamefont{K.}~\bibnamefont{Sigurdson}},
  \bibinfo{author}{\bibfnamefont{M.}~\bibnamefont{Doran}},
  \bibinfo{author}{\bibfnamefont{A.}~\bibnamefont{Kurylov}},
  \bibinfo{author}{\bibfnamefont{R.~R.} \bibnamefont{Caldwell}},
  \bibnamefont{and}
  \bibinfo{author}{\bibfnamefont{M.}~\bibnamefont{Kamionkowski}},
  \bibinfo{journal}{Phys. Rev.} \textbf{\bibinfo{volume}{D70}},
  \bibinfo{pages}{083501} (\bibinfo{year}{2004}), \bibinfo{note}{[Erratum:
  Phys. Rev.D73,089903(2006)]}, \eprint{astro-ph/0406355}.

\bibitem[{\citenamefont{Lesgourgues}(2011)}]{lesgourgues2011cosmic}
\bibinfo{author}{\bibfnamefont{J.}~\bibnamefont{Lesgourgues}},
  \bibinfo{journal}{arXiv preprint arXiv:1104.2932}  (\bibinfo{year}{2011}).

\bibitem[{\citenamefont{Padmanabhan and Finkbeiner}(2005)}]{Padmanabhan:2005es}
\bibinfo{author}{\bibfnamefont{N.}~\bibnamefont{Padmanabhan}} \bibnamefont{and}
  \bibinfo{author}{\bibfnamefont{D.~P.} \bibnamefont{Finkbeiner}},
  \bibinfo{journal}{Phys. Rev.} \textbf{\bibinfo{volume}{D72}},
  \bibinfo{pages}{023508} (\bibinfo{year}{2005}), \eprint{astro-ph/0503486}.

\bibitem[{\citenamefont{Galli et~al.}(2009)\citenamefont{Galli, Iocco, Bertone,
  and Melchiorri}}]{Galli:2009zc}
\bibinfo{author}{\bibfnamefont{S.}~\bibnamefont{Galli}},
  \bibinfo{author}{\bibfnamefont{F.}~\bibnamefont{Iocco}},
  \bibinfo{author}{\bibfnamefont{G.}~\bibnamefont{Bertone}}, \bibnamefont{and}
  \bibinfo{author}{\bibfnamefont{A.}~\bibnamefont{Melchiorri}},
  \bibinfo{journal}{Phys. Rev.} \textbf{\bibinfo{volume}{D80}},
  \bibinfo{pages}{023505} (\bibinfo{year}{2009}), \eprint{0905.0003}.

\bibitem[{\citenamefont{Slatyer et~al.}(2009)\citenamefont{Slatyer,
  Padmanabhan, and Finkbeiner}}]{Slatyer:2009yq}
\bibinfo{author}{\bibfnamefont{T.~R.} \bibnamefont{Slatyer}},
  \bibinfo{author}{\bibfnamefont{N.}~\bibnamefont{Padmanabhan}},
  \bibnamefont{and} \bibinfo{author}{\bibfnamefont{D.~P.}
  \bibnamefont{Finkbeiner}}, \bibinfo{journal}{Phys. Rev.}
  \textbf{\bibinfo{volume}{D80}}, \bibinfo{pages}{043526}
  (\bibinfo{year}{2009}), \eprint{0906.1197}.

\bibitem[{\citenamefont{Slatyer}(2016{\natexlab{a}})}]{Slatyer:2015jla}
\bibinfo{author}{\bibfnamefont{T.~R.} \bibnamefont{Slatyer}},
  \bibinfo{journal}{Phys. Rev.} \textbf{\bibinfo{volume}{D93}},
  \bibinfo{pages}{023527} (\bibinfo{year}{2016}{\natexlab{a}}),
  \eprint{1506.03811}.

\bibitem[{\citenamefont{Slatyer}(2016{\natexlab{b}})}]{Slatyer:2015kla}
\bibinfo{author}{\bibfnamefont{T.~R.} \bibnamefont{Slatyer}},
  \bibinfo{journal}{Phys. Rev.} \textbf{\bibinfo{volume}{D93}},
  \bibinfo{pages}{023521} (\bibinfo{year}{2016}{\natexlab{b}}),
  \eprint{1506.03812}.

\bibitem[{\citenamefont{Chen and Kamionkowski}(2004)}]{Chen:2003gz}
\bibinfo{author}{\bibfnamefont{X.-L.} \bibnamefont{Chen}} \bibnamefont{and}
  \bibinfo{author}{\bibfnamefont{M.}~\bibnamefont{Kamionkowski}},
  \bibinfo{journal}{Phys. Rev.} \textbf{\bibinfo{volume}{D70}},
  \bibinfo{pages}{043502} (\bibinfo{year}{2004}), \eprint{astro-ph/0310473}.

\bibitem[{\citenamefont{Slatyer and Wu}(2017)}]{Slatyer:2016qyl}
\bibinfo{author}{\bibfnamefont{T.~R.} \bibnamefont{Slatyer}} \bibnamefont{and}
  \bibinfo{author}{\bibfnamefont{C.-L.} \bibnamefont{Wu}},
  \bibinfo{journal}{Phys. Rev.} \textbf{\bibinfo{volume}{D95}},
  \bibinfo{pages}{023010} (\bibinfo{year}{2017}), \eprint{1610.06933}.

\bibitem[{\citenamefont{Poulin et~al.}(2017)\citenamefont{Poulin, Lesgourgues,
  and Serpico}}]{Poulin:2016anj}
\bibinfo{author}{\bibfnamefont{V.}~\bibnamefont{Poulin}},
  \bibinfo{author}{\bibfnamefont{J.}~\bibnamefont{Lesgourgues}},
  \bibnamefont{and} \bibinfo{author}{\bibfnamefont{P.~D.}
  \bibnamefont{Serpico}}, \bibinfo{journal}{JCAP}
  \textbf{\bibinfo{volume}{1703}}, \bibinfo{pages}{043} (\bibinfo{year}{2017}),
  \eprint{1610.10051}.

\bibitem[{\citenamefont{McDermott et~al.}(2011)\citenamefont{McDermott, Yu, and
  Zurek}}]{McDermott:2010pa}
\bibinfo{author}{\bibfnamefont{S.~D.} \bibnamefont{McDermott}},
  \bibinfo{author}{\bibfnamefont{H.-B.} \bibnamefont{Yu}}, \bibnamefont{and}
  \bibinfo{author}{\bibfnamefont{K.~M.} \bibnamefont{Zurek}},
  \bibinfo{journal}{Phys. Rev.} \textbf{\bibinfo{volume}{D83}},
  \bibinfo{pages}{063509} (\bibinfo{year}{2011}), \eprint{1011.2907}.

\bibitem[{\citenamefont{Ade et~al.}(2015)}]{Planck:2015xua}
\bibinfo{author}{\bibfnamefont{P.}~\bibnamefont{Ade}} \bibnamefont{et~al.}
  (\bibinfo{collaboration}{Planck}) (\bibinfo{year}{2015}),
  \eprint{1502.01589}.

\bibitem[{\citenamefont{{Tseliakhovich} and
  {Hirata}}(2010)}]{2010PhRvD..82h3520T}
\bibinfo{author}{\bibfnamefont{D.}~\bibnamefont{{Tseliakhovich}}}
  \bibnamefont{and} \bibinfo{author}{\bibfnamefont{C.}~\bibnamefont{{Hirata}}},
  \bibinfo{journal}{\prd} \textbf{\bibinfo{volume}{82}}, \bibinfo{eid}{083520}
  (\bibinfo{year}{2010}), \eprint{1005.2416}.

\bibitem[{\citenamefont{{Audren} et~al.}(2013)\citenamefont{{Audren},
  {Lesgourgues}, {Benabed}, and {Prunet}}}]{audren7183conservative}
\bibinfo{author}{\bibfnamefont{B.}~\bibnamefont{{Audren}}},
  \bibinfo{author}{\bibfnamefont{J.}~\bibnamefont{{Lesgourgues}}},
  \bibinfo{author}{\bibfnamefont{K.}~\bibnamefont{{Benabed}}},
  \bibnamefont{and} \bibinfo{author}{\bibfnamefont{S.}~\bibnamefont{{Prunet}}},
  \bibinfo{journal}{JCAP} \textbf{\bibinfo{volume}{2}}, \bibinfo{eid}{001}
  (\bibinfo{year}{2013}), \eprint{1210.7183}.

\bibitem[{\citenamefont{Aghanim et~al.}(2016)}]{Aghanim:2015xee}
\bibinfo{author}{\bibfnamefont{N.}~\bibnamefont{Aghanim}} \bibnamefont{et~al.}
  (\bibinfo{collaboration}{Planck}), \bibinfo{journal}{Astron. Astrophys.}
  \textbf{\bibinfo{volume}{594}}, \bibinfo{pages}{A11} (\bibinfo{year}{2016}),
  \eprint{1507.02704}.

\bibitem[{\citenamefont{Das et~al.}(2014)}]{Das:2013zf}
\bibinfo{author}{\bibfnamefont{S.}~\bibnamefont{Das}} \bibnamefont{et~al.},
  \bibinfo{journal}{JCAP} \textbf{\bibinfo{volume}{1404}}, \bibinfo{pages}{014}
  (\bibinfo{year}{2014}), \eprint{1301.1037}.

\bibitem[{\citenamefont{Dunkley et~al.}(2013)}]{Dunkley:2013vu}
\bibinfo{author}{\bibfnamefont{J.}~\bibnamefont{Dunkley}} \bibnamefont{et~al.},
  \bibinfo{journal}{JCAP} \textbf{\bibinfo{volume}{1307}}, \bibinfo{pages}{025}
  (\bibinfo{year}{2013}), \eprint{1301.0776}.

\bibitem[{\citenamefont{Calabrese et~al.}(2013)}]{Calabrese:2013jyk}
\bibinfo{author}{\bibfnamefont{E.}~\bibnamefont{Calabrese}}
  \bibnamefont{et~al.}, \bibinfo{journal}{Phys. Rev.}
  \textbf{\bibinfo{volume}{D87}}, \bibinfo{pages}{103012}
  (\bibinfo{year}{2013}), \eprint{1302.1841}.

\bibitem[{\citenamefont{Finkbeiner et~al.}(2012)\citenamefont{Finkbeiner,
  Galli, Lin, and Slatyer}}]{Finkbeiner:2011dx}
\bibinfo{author}{\bibfnamefont{D.~P.} \bibnamefont{Finkbeiner}},
  \bibinfo{author}{\bibfnamefont{S.}~\bibnamefont{Galli}},
  \bibinfo{author}{\bibfnamefont{T.}~\bibnamefont{Lin}}, \bibnamefont{and}
  \bibinfo{author}{\bibfnamefont{T.~R.} \bibnamefont{Slatyer}},
  \bibinfo{journal}{Phys.Rev.} \textbf{\bibinfo{volume}{D85}},
  \bibinfo{pages}{043522} (\bibinfo{year}{2012}), \eprint{1109.6322}.

\bibitem[{\citenamefont{{Jungman} et~al.}(1996)\citenamefont{{Jungman},
  {Kamionkowski}, {Kosowsky}, and {Spergel}}}]{jungman96b}
\bibinfo{author}{\bibfnamefont{G.}~\bibnamefont{{Jungman}}},
  \bibinfo{author}{\bibfnamefont{M.}~\bibnamefont{{Kamionkowski}}},
  \bibinfo{author}{\bibfnamefont{A.}~\bibnamefont{{Kosowsky}}},
  \bibnamefont{and} \bibinfo{author}{\bibfnamefont{D.~N.}
  \bibnamefont{{Spergel}}}, \bibinfo{journal}{\prd}
  \textbf{\bibinfo{volume}{54}}, \bibinfo{pages}{1332} (\bibinfo{year}{1996}),
  \eprint{arXiv:astro-ph/9512139}.

\bibitem[{\citenamefont{{Tegmark} et~al.}(1997)\citenamefont{{Tegmark},
  {Taylor}, and {Heavens}}}]{tegmark97}
\bibinfo{author}{\bibfnamefont{M.}~\bibnamefont{{Tegmark}}},
  \bibinfo{author}{\bibfnamefont{A.~N.} \bibnamefont{{Taylor}}},
  \bibnamefont{and} \bibinfo{author}{\bibfnamefont{A.~F.}
  \bibnamefont{{Heavens}}}, \bibinfo{journal}{\apj}
  \textbf{\bibinfo{volume}{480}}, \bibinfo{pages}{22} (\bibinfo{year}{1997}),
  \eprint{arXiv:astro-ph/9603021}.

\bibitem[{\citenamefont{Verde}(2010)}]{Verde:2009tu}
\bibinfo{author}{\bibfnamefont{L.}~\bibnamefont{Verde}},
  \bibinfo{journal}{Lect. Notes Phys.} \textbf{\bibinfo{volume}{800}},
  \bibinfo{pages}{147} (\bibinfo{year}{2010}), \eprint{0911.3105}.

\bibitem[{\citenamefont{Furlanetto et~al.}(2006)\citenamefont{Furlanetto, Oh,
  and Briggs}}]{Furlanetto:2006jb}
\bibinfo{author}{\bibfnamefont{S.}~\bibnamefont{Furlanetto}},
  \bibinfo{author}{\bibfnamefont{S.~P.} \bibnamefont{Oh}}, \bibnamefont{and}
  \bibinfo{author}{\bibfnamefont{F.}~\bibnamefont{Briggs}},
  \bibinfo{journal}{Phys. Rept.} \textbf{\bibinfo{volume}{433}},
  \bibinfo{pages}{181} (\bibinfo{year}{2006}), \eprint{astro-ph/0608032}.

\bibitem[{\citenamefont{Tashiro et~al.}(2014)\citenamefont{Tashiro, Kadota, and
  Silk}}]{Tashiro:2014tsa}
\bibinfo{author}{\bibfnamefont{H.}~\bibnamefont{Tashiro}},
  \bibinfo{author}{\bibfnamefont{K.}~\bibnamefont{Kadota}}, \bibnamefont{and}
  \bibinfo{author}{\bibfnamefont{J.}~\bibnamefont{Silk}},
  \bibinfo{journal}{Phys. Rev.} \textbf{\bibinfo{volume}{D90}},
  \bibinfo{pages}{083522} (\bibinfo{year}{2014}), \eprint{1408.2571}.

\bibitem[{\citenamefont{Muñoz et~al.}(2015)\citenamefont{Muñoz, Kovetz, and
  Ali-Haïmoud}}]{Munoz:2015bca}
\bibinfo{author}{\bibfnamefont{J.~B.} \bibnamefont{Muñoz}},
  \bibinfo{author}{\bibfnamefont{E.~D.} \bibnamefont{Kovetz}},
  \bibnamefont{and}
  \bibinfo{author}{\bibfnamefont{Y.}~\bibnamefont{Ali-Haïmoud}},
  \bibinfo{journal}{Phys. Rev.} \textbf{\bibinfo{volume}{D92}},
  \bibinfo{pages}{083528} (\bibinfo{year}{2015}), \eprint{1509.00029}.

\bibitem[{\citenamefont{Ali-Haïmoud et~al.}(2014)\citenamefont{Ali-Haïmoud,
  Meerburg, and Yuan}}]{Ali-Haimoud:2013hpa}
\bibinfo{author}{\bibfnamefont{Y.}~\bibnamefont{Ali-Haïmoud}},
  \bibinfo{author}{\bibfnamefont{P.~D.} \bibnamefont{Meerburg}},
  \bibnamefont{and} \bibinfo{author}{\bibfnamefont{S.}~\bibnamefont{Yuan}},
  \bibinfo{journal}{Phys. Rev.} \textbf{\bibinfo{volume}{D89}},
  \bibinfo{pages}{083506} (\bibinfo{year}{2014}), \eprint{1312.4948}.

\bibitem[{\citenamefont{Muñoz and Loeb}(2017)}]{Munoz:2017qpy}
\bibinfo{author}{\bibfnamefont{J.~B.} \bibnamefont{Muñoz}} \bibnamefont{and}
  \bibinfo{author}{\bibfnamefont{A.}~\bibnamefont{Loeb}},
  \bibinfo{journal}{JCAP} \textbf{\bibinfo{volume}{1711}}, \bibinfo{pages}{043}
  (\bibinfo{year}{2017}), \eprint{1708.08923}.

\bibitem[{\citenamefont{Hui and Gnedin}(1997)}]{Hui:1997dp}
\bibinfo{author}{\bibfnamefont{L.}~\bibnamefont{Hui}} \bibnamefont{and}
  \bibinfo{author}{\bibfnamefont{N.~Y.} \bibnamefont{Gnedin}},
  \bibinfo{journal}{Mon. Not. Roy. Astron. Soc.}
  \textbf{\bibinfo{volume}{292}}, \bibinfo{pages}{27} (\bibinfo{year}{1997}),
  \eprint{astro-ph/9612232}.

\bibitem[{\citenamefont{Sanderbeck et~al.}(2016)\citenamefont{Sanderbeck,
  D'Aloisio, and McQuinn}}]{Sanderbeck:2015bba}
\bibinfo{author}{\bibfnamefont{P.~R.~U.} \bibnamefont{Sanderbeck}},
  \bibinfo{author}{\bibfnamefont{A.}~\bibnamefont{D'Aloisio}},
  \bibnamefont{and} \bibinfo{author}{\bibfnamefont{M.~J.}
  \bibnamefont{McQuinn}}, \bibinfo{journal}{Mon. Not. Roy. Astron. Soc.}
  \textbf{\bibinfo{volume}{460}}, \bibinfo{pages}{1885} (\bibinfo{year}{2016}),
  \eprint{1511.05992}.

\bibitem[{\citenamefont{Viel et~al.}(2013)\citenamefont{Viel, Becker, Bolton,
  and Haehnelt}}]{Viel:2013apy}
\bibinfo{author}{\bibfnamefont{M.}~\bibnamefont{Viel}},
  \bibinfo{author}{\bibfnamefont{G.~D.} \bibnamefont{Becker}},
  \bibinfo{author}{\bibfnamefont{J.~S.} \bibnamefont{Bolton}},
  \bibnamefont{and} \bibinfo{author}{\bibfnamefont{M.~G.}
  \bibnamefont{Haehnelt}}, \bibinfo{journal}{Phys. Rev.}
  \textbf{\bibinfo{volume}{D88}}, \bibinfo{pages}{043502}
  (\bibinfo{year}{2013}), \eprint{1306.2314}.

\bibitem[{\citenamefont{McDonald et~al.}(2005)}]{McDonald:2004xn}
\bibinfo{author}{\bibfnamefont{P.}~\bibnamefont{McDonald}} \bibnamefont{et~al.}
  (\bibinfo{collaboration}{SDSS}), \bibinfo{journal}{Astrophys. J.}
  \textbf{\bibinfo{volume}{635}}, \bibinfo{pages}{761} (\bibinfo{year}{2005}),
  \eprint{astro-ph/0407377}.

\bibitem[{\citenamefont{Hu and Silk}(1993{\natexlab{a}})}]{Hu:1992dc}
\bibinfo{author}{\bibfnamefont{W.}~\bibnamefont{Hu}} \bibnamefont{and}
  \bibinfo{author}{\bibfnamefont{J.}~\bibnamefont{Silk}},
  \bibinfo{journal}{Phys. Rev.} \textbf{\bibinfo{volume}{D48}},
  \bibinfo{pages}{485} (\bibinfo{year}{1993}{\natexlab{a}}).

\bibitem[{\citenamefont{Chluba}(2013)}]{Chluba:2013wsa}
\bibinfo{author}{\bibfnamefont{J.}~\bibnamefont{Chluba}},
  \bibinfo{journal}{Mon.Not.Roy.Astron.Soc.} \textbf{\bibinfo{volume}{436}},
  \bibinfo{pages}{2232} (\bibinfo{year}{2013}), \eprint{1304.6121}.

\bibitem[{\citenamefont{Hu and Silk}(1993{\natexlab{b}})}]{Hu:1993gc}
\bibinfo{author}{\bibfnamefont{W.}~\bibnamefont{Hu}} \bibnamefont{and}
  \bibinfo{author}{\bibfnamefont{J.}~\bibnamefont{Silk}},
  \bibinfo{journal}{Phys. Rev. Lett.} \textbf{\bibinfo{volume}{70}},
  \bibinfo{pages}{2661} (\bibinfo{year}{1993}{\natexlab{b}}).

\bibitem[{\citenamefont{Hill et~al.}(2015)\citenamefont{Hill, Battaglia,
  Chluba, Ferraro, Schaan, and Spergel}}]{Hill:2015tqa}
\bibinfo{author}{\bibfnamefont{J.~C.} \bibnamefont{Hill}},
  \bibinfo{author}{\bibfnamefont{N.}~\bibnamefont{Battaglia}},
  \bibinfo{author}{\bibfnamefont{J.}~\bibnamefont{Chluba}},
  \bibinfo{author}{\bibfnamefont{S.}~\bibnamefont{Ferraro}},
  \bibinfo{author}{\bibfnamefont{E.}~\bibnamefont{Schaan}}, \bibnamefont{and}
  \bibinfo{author}{\bibfnamefont{D.~N.} \bibnamefont{Spergel}},
  \bibinfo{journal}{Phys. Rev. Lett.} \textbf{\bibinfo{volume}{115}},
  \bibinfo{pages}{261301} (\bibinfo{year}{2015}), \eprint{1507.01583}.

\bibitem[{\citenamefont{Fixsen et~al.}(1996)\citenamefont{Fixsen, Cheng, Gales,
  Mather, Shafer et~al.}}]{Fixsen:1996nj}
\bibinfo{author}{\bibfnamefont{D.}~\bibnamefont{Fixsen}},
  \bibinfo{author}{\bibfnamefont{E.}~\bibnamefont{Cheng}},
  \bibinfo{author}{\bibfnamefont{J.}~\bibnamefont{Gales}},
  \bibinfo{author}{\bibfnamefont{J.~C.} \bibnamefont{Mather}},
  \bibinfo{author}{\bibfnamefont{R.}~\bibnamefont{Shafer}},
  \bibnamefont{et~al.}, \bibinfo{journal}{Astrophys.J.}
  \textbf{\bibinfo{volume}{473}}, \bibinfo{pages}{576} (\bibinfo{year}{1996}),
  \eprint{astro-ph/9605054}.

\bibitem[{\citenamefont{Kogut et~al.}(2011)}]{Kogut:2011xw}
\bibinfo{author}{\bibfnamefont{A.}~\bibnamefont{Kogut}} \bibnamefont{et~al.},
  \bibinfo{journal}{JCAP} \textbf{\bibinfo{volume}{1107}}, \bibinfo{pages}{025}
  (\bibinfo{year}{2011}), \eprint{1105.2044}.

\bibitem[{\citenamefont{Dolgov et~al.}(2013)\citenamefont{Dolgov, Dubovsky,
  Rubtsov, and Tkachev}}]{Dolgov:2013una}
\bibinfo{author}{\bibfnamefont{A.~D.} \bibnamefont{Dolgov}},
  \bibinfo{author}{\bibfnamefont{S.~L.} \bibnamefont{Dubovsky}},
  \bibinfo{author}{\bibfnamefont{G.~I.} \bibnamefont{Rubtsov}},
  \bibnamefont{and} \bibinfo{author}{\bibfnamefont{I.~I.}
  \bibnamefont{Tkachev}}, \bibinfo{journal}{Phys. Rev.}
  \textbf{\bibinfo{volume}{D88}}, \bibinfo{pages}{117701}
  (\bibinfo{year}{2013}), \eprint{1310.2376}.

\end{thebibliography}

\end{document}